\documentclass[%
 reprint,
 superscriptaddress,
 amsmath,
 amssymb,
 aps,
prb,
 tightenlines,
 onecolumn,
]{revtex4-2}

\usepackage{color}
\usepackage{amsmath}
\usepackage{graphicx}
\usepackage{dcolumn}
\usepackage{bm}
\usepackage{booktabs}
\usepackage{multirow}

\usepackage{natbib}

\begin{document}


\title{
Topological magnon insulators in two-dimensional van der Waals ferromagnets CrSiTe$_3$ and CrGeTe$_3$: towards intrinsic gap-tunability
}

\author{Fengfeng Zhu}
\thanks{These authors contributed equally to this work.}%
\email{Corresponding author. E-mail: f.zhu@fz-juelich.de}
\affiliation{J\"ulich Centre for Neutron Science (JCNS) at Heinz Maier-Leibnitz Zentrum (MLZ), Forschungszentrum J\"ulich, Lichtenbergstrasse 1, D-85747 Garching, Germany}
\affiliation{Department of Physics and Astronomy, Shanghai Jiao Tong University, 200240 Shanghai, China}

\author{Lichuan Zhang}\thanks{These authors contributed equally to this work.}
\affiliation{Peter Gr\"unberg Institut and Institute for Advanced Simulation,
Forschungszentrum J\"ulich and JARA, 52425 J\"ulich, Germany}
\affiliation{Department of Physics, RWTH Aachen University, 52056 Aachen, Germany}

\author{Xiao Wang}\thanks{These authors contributed equally to this work.}
\affiliation{J\"ulich Centre for Neutron Science (JCNS) at Heinz Maier-Leibnitz Zentrum (MLZ), Forschungszentrum J\"ulich, Lichtenbergstrasse 1, D-85747 Garching, Germany}

\author{Flaviano Jos\'e dos Santos}
\affiliation{Peter Gr\"unberg Institut and Institute for Advanced Simulation,
Forschungszentrum J\"ulich and JARA, 52425 J\"ulich, Germany}
\affiliation{Theory and Simulation of Materials (THEOS), and National Centre for Computational Design and Discovery of Novel Materials (MARVEL), \'Ecole Polytechnique F\'ed\'erale de Lausanne, 1015 Lausanne, Switzerland}

\author{Junda Song}
\affiliation{J\"ulich Centre for Neutron Science (JCNS) at Heinz Maier-Leibnitz Zentrum (MLZ), Forschungszentrum J\"ulich, Lichtenbergstrasse 1, D-85747 Garching, Germany}

\author{Thomas Mueller}
\affiliation{J\"ulich Centre for Neutron Science (JCNS) at Heinz Maier-Leibnitz Zentrum (MLZ), Forschungszentrum J\"ulich, Lichtenbergstrasse 1, D-85747 Garching, Germany}

\author{Karin Schmalzl}
\affiliation{J\"ulich Centre for Neutron Science (JCNS) at ILL, Forschungszentrum J\"ulich, F-38000 Grenoble, France}

\author{Wolfgang F. Schmidt}
\affiliation{J\"ulich Centre for Neutron Science (JCNS) at ILL, Forschungszentrum J\"ulich, F-38000 Grenoble, France}

\author{Alexandre Ivanov}
\affiliation{Institut Laue-Langevin, 71 Avenue des Martyrs CS 20156, 38042 Grenoble Cedex 9, France}

\author{Jitae T. Park}
\affiliation{Heinz Maier-Leibnitz Zentrum (MLZ),
Technische Universität München, D-85747 Garching, Germany}

\author{Jianhui Xu}
\affiliation{Helmholtz-Zentrum Berlin für Materialien und Energie GmbH, Hahn-Meitner-Platz 1, D-14109 Berlin, Germany}
\affiliation{Heinz Maier-Leibnitz Zentrum (MLZ),
Technische Universität München, D-85747 Garching, Germany}

\author{Jie Ma}
\affiliation{Department of Physics and Astronomy, Shanghai Jiao Tong University, 200240 Shanghai, China}

\author{Samir Lounis}
\affiliation{Peter Gr\"unberg Institut and Institute for Advanced Simulation,
Forschungszentrum J\"ulich and JARA, 52425 J\"ulich, Germany}
\affiliation{Faculty of Physics, University of Duisburg-Essen and CENIDE, 47053 Duisburg, Germany}

\author{Stefan Bl\"ugel}
\affiliation{Peter Gr\"unberg Institut and Institute for Advanced Simulation,
Forschungszentrum J\"ulich and JARA, 52425 J\"ulich, Germany}

\author{Yuriy Mokrousov}
\email{Corresponding author. E-mail: y.mokrousov@fz-juelich.de}
\affiliation{Peter Gr\"unberg Institut and Institute for Advanced Simulation,
Forschungszentrum J\"ulich and JARA, 52425 J\"ulich, Germany}
\affiliation{Institute of Physics, Johannes Gutenberg University Mainz, 55099 Mainz, Germany}

\author{Yixi Su}
\email{Corresponding author. E-mail: y.su@fz-juelich.de}
\affiliation{J\"ulich Centre for Neutron Science (JCNS) at Heinz Maier-Leibnitz Zentrum (MLZ), Forschungszentrum J\"ulich, Lichtenbergstrasse 1, D-85747 Garching, Germany}

\author{Thomas Br\"uckel}
\affiliation{J\"ulich Centre for Neutron Science JCNS and Peter Grünberg Institut PGI, JARA-FIT, Forschungszentrum J\"ulich, D-52425 J\"ulich, Germany}


\begin{abstract}
The bosonic analogues of topological insulators have been proposed in numerous theoretical works, but their experimental realization is still very rare, especially for spin systems. Recently,  two-dimensional (2D) honeycomb van der Waals (vdW) ferromagnets  have  emerged as a new platform for topological spin excitations.
Here, via a comprehensive inelastic neutron scattering study and theoretical analysis of the spin-wave excitations, we report the realization of topological magnon insulators in CrXTe$_3$ (X=Si, Ge) compounds.
The nontrivial nature and intrinsic tunability of the gap opening at the magnon band-crossing Dirac points are confirmed, while the emergence of the corresponding in-gap topological edge states is demonstrated theoretically.
The realization of topological magnon insulators with intrinsic gap-tunability in this class of remarkable 2D materials will undoubtedly lead to new and fascinating technological applications in the domain of magnonics and topological spintronics.
\end{abstract}

\maketitle


\section{Introduction}
Recent theoretical predictions and experimental realizations of exotic quasi-particles and topological excitations in condensed matter have led to a tremendous research interest in topological quantum materials \cite{Haldane1988,Kane2005,Hasan2010,Qi2011a,Armitage2018a}.
The topology of the electronic band structure is closely linked to the Berry curvature in the $k$-space \cite{Nagaosa2010,Xiao2010}, for which the spin-orbit coupling (SOC) plays a key role.
In principle, it is also possible to realize non-trivial topology  in a bosonic system, such as a system of magnonic excitations~\cite{Chisnell2015a,Owerre2016c,Wang2017e,Chen2018,Joshi2018,Pantaleon2018,McClarty2018a,Bao2018,Yao2018,Yuan2020,Aguilera2020}, since the band topology can be treated independently from the statistical nature of the particles.
Such topological excitations and the corresponding non-trivial in-gap edge states are chiral and robust against disorder,
it is thus believed that the emergence and manipulation of the topological magnonic states bares a tremendous promise for future applications in magnonics and topological spintronics, such as~e.g.~quantized pumping of magnons~\cite{Mei2019},
spin-wave beam splitter~\cite{Wang2017e}, magnon waveguides~\cite{Mook2015}, chiral travelling-wave magnon amplifiers~\cite{Malz2019} and magnon-driven orbitronics~\cite{Zhang2020d}.

However, in contrast to the case of fermionic systems, the topological excitations have been realized only in very few bosonic systems. This is largely owing to the lack of suitable candidate materials as well as to the tremendous challenges in experimental probes of uncharged bosonic topological excitations, and, in particular, in a direct detection of in-gap surface/edge states. For instance, inelastic neutron scattering may be the method of choice~e.g.~for disentangling topological excitation gaps, but, to directly probe the in-gap edge excitation mode is unfortunately beyond its reach even with the most powerful neutron instruments available nowadays. Nevertheless, since topological magnon edge states contribute to the transverse thermal Hall conductivity, it has recently been demonstrated that the nontrivial topological nature of the magnonic bands can be manifested experimentally by the thermal Hall effect even in a charge-neutral spin system \cite{Katsura2010,Onose2010a,Matsumoto2011a,Owerre2016b}.
Following the pioneering study of the magnon thermal Hall effect in the pyrochlore ferromagnet Lu$_2$V$_2$O$_7$ \cite{Katsura2010,Onose2010a}, the evidence for the gapped magnonic topological insulating phase was reported in the Ising-like kagome ferromagnet Cu[1,3-bdc] under applied magnetic field  via inelastic neutron scattering \cite{Chisnell2015a}.
The search for topological excitations in various spin systems has been intensified since then.
Recently, Dirac magnons that exhibit symmetry-protected band crossings have been observed in a three-dimensional (3D) Heisenberg antiferromagnet Cu$_3$TeO$_6$ \cite{Bao2018,Yao2018} and in a 3D quantum $XY$ magnet CoTiO$_3$ \cite{Yuan2020}.
Meanwhile, 2D honeycomb vdW magnets~\cite{Gong2017,Ding2020,Kasahara2018a}, such as the 2D ferromagnets CrGeTe$_3$ and Fe$_3$GeTe$_2$ as well as the Kitaev material $\alpha$-RuCl$_3$, have  attracted tremendous research interest owing to their fascinating physical properties. Since the recent observation of the gapped topological excitations in ferromagnetic CrI$_3$ \cite{Chen2018}, 2D vdW honeycomb-lattice magnets start to emerge as a unique platform for the exploration of the topological magnonics~\cite{Owerre2016c,Joshi2018,McClarty2018a}.
In this class of materials, as schematically shown in Fig.1(a), a topological gap can be opened at the magnon band-crossing Dirac points when a sufficiently large SOC is present. Pronounced SOC would in turn give rise to antisymmetric or anisotropic exchange interactions $-$ such as the Dzyaloshinskii-Moriya (DM) \cite{Moriya1960} and Kitaev interactions $-$ which can drive non-trivial magnonic topology.

While the current research focus of the community falls primarily on few-layer vdW heterostructures and Mori\'e superlattices, a variety of recently discovered  quantum phenomena in bulk vdW materials $-$  e.g. the half-integer thermal quantum Hall effect and fractionalized Majorana fermions in $\alpha$-RuCl$_3$ \cite{Kasahara2018a} or topological magnetic textures in Fe$_3$GeTe$_2$ \cite{Ding2020} $-$ clearly indicate the potential of bulk 2D vdW magnetic materials as a rich playground in topological quantum matter research.
A thorough understanding of intrinsic magnetic interactions and emergent topological properties in single-crystal bulk samples will  undoubtedly  stimulate further research on monolayer, few-layer and Mori\'e-based counterparts.
In this context, 2D vdW CrXTe$_3$ compounds emerge as one of the very promising representatives in this class. 
In fact, one  would expect the existence of topological magnons in CrXTe$_3$ similarly to CrI$_3$~\cite{Owerre2016c,Owerre2016b,Aguilera2020}, as long as SOC and the antisymmetric exchange interactions are strong enough in this ferromagnetic honeycomb system.
Furthermore, due to high versatility in terms of chemical composition, a range of topological magnon insulators with intrinsic gap-tunability may be realized in CrXTe$_3$ family.
As such, the intrinsic control of topological magnon properties, for instance, via the engineering of DM interaction, may lead to novel applications in magnonic spintronics and quantum technology.

In this work, we present a comprehensive inelastic neutron scattering study of the spin-wave excitations in single-crystal CrSiTe$_3$ and CrGeTe$_3$, which belong to a family of 2D vdW honeycomb ferromagnets.
Our inelastic neutron scattering experiments performed at low temperatures show clear dispersive magnonic bands and a well-resolved band gap opening at the high-symmetry $K$ points in the Brillouin zone (BZ).
By fitting to experimental data within the linear spin-wave theory (LSWT), the magnitude of  exchange interactions in studied materials 
has been determined.
Moreover, the observed band gap opening was ascribed to the antisymmetric exchange interactions, namely the DM interaction, and a spin Hamiltonian model 
including the second-nearest-neighbor (2nd-NN) DM interaction could provide a very good description of the magnonic dispersion in CrXTe$_3$. 
In line with expectations, the size of the topological gap was found to be strongly dependent on the strength of the DM interaction that intrinsically originates from SOC in this system.
Furthermore, the Chern numbers of the magnonic bands were found to be non-zero, thus indicating that the band gap opening is indeed topologically nontrivial and corresponding edge states could emerge inside the gap.
By assessing theoretically the magnitude of the  thermal Hall effect 
we find that the topologically nontrivial features could be detected in thermal transport measurements in two considered systems. 
Based on the compelling evidence obtained in our inelastic neutron scattering experiments and theoretical calculations, we thus conclude that the exotic topological magnon insulator, that is intrinsically gap-tunable, can be ideally realized in the family of 2D vdW honeycomb ferromagnets CrXTe$_3$.

\section{Results and Discussion}
High-quality single crystals of CrXTe$_3$ (X=Si, Ge) were grown by the flux method, and were carefully examined by X-ray Laue, single-crystal X-ray and neutron diffraction.
The Cr$^{3+}$ ions with $S=3/2$ in CrXTe$_3$ (X=Si, Ge) occupy an $ABC$ stacked honeycomb lattice, as shown in Fig.1(b).
Despite the presence of the Si/Ge atoms, each layer shares a similar atomic structure to that of the chromium trihalides CrI$_3$ \cite{McGuire2015} and CrBr$_3$ at low temperature.
All the Cr$^{3+}$ ions are located in the center of the edge-shared trigonal distorted octahedra (D$_{3d}$ symmetry) composed of Te atoms, as shown in Fig.1(c).
Below $T_c$, both materials exhibit ferromagnetic order with the magnetic moments aligned along the $c$ axis.
As shown in Fig.S1, the magnetic susceptibilities measured with fields applied along the $a$ and $c$ axes clearly show ferromagnetic phase transitions occurring at $\sim$33 K and $\sim$63 K for CrSiTe$_3$ and CrGeTe$_3$ respectively, with the easy axis  along the $c$ axis.
As the isothermal magnetization curves shown in Fig.S1, their magnetic moments can be easily aligned and saturated along both the $a$ and $c$ axes by an application of a small magnetic field, which indicates a small magnetic anisotropy energy in these two materials, consistent with the previous magnetization measurements \cite{Casto2015,Liu2016d,Kim2019,Zeisner2019}.
In addition, the ferromagnetic transitions are further confirmed by neutron diffraction and the transition temperatures $T_c$ are extracted from the fitting of the temperature dependent intensity of the (1,1,0) magnetic Bragg peak.
Among them, the magnetic moment direction of CrGeTe$_3$ in the ferromagnetic phase is determined along the $c$ axis by polarized neutron diffraction, as shown in Fig.S2.

In Fig.2, we show the magnon spectra of CrSiTe$_3$ along the high-symmetry directions [-1,H,0] and [H,H,3] in the (H,K,0) reciprocal plane as well as along the [1,1,L] direction in the (H,H,L) reciprocal plane.
The measured momentum transfer $Q$ positions in reciprocal space are shown as the black solid lines in the inset of Fig.2(c).
The measured magnon bands lie below 16 meV and show clear dispersion along all directions.
As shown in Fig.2(a), the low-energy magnon mode (referred to as ``acoustic'' magnon mode) emerges from the BZ center $\mathit{\Gamma}$ point and reaches its maximum energy with a gap opening at the $M$ point of the in-plane BZ boundary, consistently with its ferromagnetic nature.
Along the [H,H,3] direction, in the vicinity of the $K$ points, as shown in the Fig.2(b), a small spin gap is faintly visible at $E\approx$ 9 meV.
Besides, as highlighted by the white arrow in the inset of Fig.2(b), a new branch connected to the acoustic magnon mode was observed along the BZ boundary $K$-$M$-$K^\prime$ direction although its intensity is rather weak.
From a zoom-in plot in Fig.2(g), the high-energy branch (referred to as ``optical'' magnon mode) is found to be less dispersive as compared to the acoustic magnon mode, and it also seems disconnected from the acoustic branch at $K$ points.
To further confirm whether a magnon gap is opened at the $K$ point, the line profiles of the constant-Q scans at two different $K$ points are shown in Fig.2(h), where a gap opening of about 2 meV can be clearly resolved.
Subsequently, the maximum of the acoustic and the minimum of the optical magnon mode at the $K$ point are determined as 8.6(1) meV and 10.7(1) meV by the multi-peak fitting, respectively.
Given that the magnitude of the gap is quite large and no clear phonon modes are observed, we conclude that the magnon-phonon hybridization is unlikely to be the origin of the observed gap opening.
Instead, introducing an antisymmetric exchange interaction,  such as the DM interaction, into the spin Hamiltonian could potentially induce such a considerable gap at the $K$ point.
As for the [1,1,L] direction, the acoustic magnon mode only extends up to 2 meV, which is much smaller than that in the $\mathit{\Gamma}$-$M$ and $\mathit{\Gamma}$-$K$ directions, indicating that the coupling between vdW layers is notably weak and that the magnon spectra are mainly dictated by the intra-layer exchange interactions.

As for the analogous CrGeTe$_3$, whose transition temperature $T_c$ is nearly doubled as compared to that of CrSiTe$_3$, we also measured its spin waves along all high-symmetry directions, and the black solid lines in the inset of Fig.3(b) show the corresponding $Q$ positions for the measurements.
The magnon modes in the (H,K,0) reciprocal plane are shown in Fig.3(a), where two branches, namely the acoustic and optical modes, can be easily distinguished, and the magnon modes along [1,1,L] directions, dictated by the inter-layer exchange interactions, are shown in Fig.3(b).
As expected, the acoustic mode has a stronger dispersion than the optical mode and also emerges from the magnetic BZ center $\mathit{\Gamma}$ point at low energy.
Except for the scale of the energy transfers, the overall feature of the magnon modes of CrGeTe$_3$ appears very similar to that of CrSiTe$_3$ including the gap opening at the $K$ point.
Interestingly, the acoustic magnon mode in the (H,K,0) plane reaches its energy maximum at around 15 meV which is nearly two times that of CrSiTe$_3$, but along the out-of-plane direction, the magnon mode only extends to 2.5 meV, basically the same as in CrSiTe$_3$.
It indicates that replacing the Si atoms by heavier Ge atoms in CrXTe$_3$ will significantly enhance the intra-layer exchange interactions and meanwhile will have little influence on the interlayer exchange interactions.
This suggests that the exchange interactions can be tuned by nonmagnetic atoms without destroying the magnetic network, and that the strength of SOC of non-magnetic atoms may also play an important role in the super-exchange interactions in this vdW honeycomb ferromagnet family.

Furthermore, one can notice that the optical mode of CrGeTe$_3$ is rather broad in energy even after taking into consideration the instrument resolution of $\sim$2 meV at such energy transfer, and is quite intense, in contrast to the situation of CrSiTe$_3$.
In fact, we find that the intensity of this optical magnon branch at the $K$ point is about 4 times higher than the prediction of our LSWT calculation. 
Given that the optical phonon mode of CrGeTe$_3$ has a similar energy scale \cite{Wang2019}, and that the evidence for the spin-lattice coupling in CrGeTe$_3$ has been already reported in previous Raman experiments \cite{Tian2016a,Sun2018}, we tend to believe that both the broadening in energy and the intensity enhancement could be a result of possible magnon-phonon interactions.
Interestingly, the phonon modes around 25 meV are indeed mainly contributed by the magnetic atoms Cr \cite{Wang2021}, the coexistence of the phonon and magnon modes in the same energy range may allow for a possible spectral weight transfer between them, thus leading to the observed unusual intensity enhancement.
Generally, a dynamic spin-lattice coupling can create hybridization gaps or broadening in the magnon spectra at the intersection points of the coupled magnon and phonon modes.

Regardless of the unusual broadening of the optical branches, a gap opening at the $K$ point is still resolvable.
In Fig.3(e)-(g), three line profiles of energy scans are extracted from the vicinity of the $K$ point and fitted by two Gauss peaks to accurately determine the size of the opened gap, and the corresponding positions of momentum transfer in reciprocal space in the measurements are marked by the black circles in Fig.(h)-(j).
From Fig3.(f), a weak but clear shoulder can be seen around 15 meV which denotes the band maximum of the acoustic mode, and subsequently the size of the opened gap at the $K$ point can be determined as $\sim$5 meV from the fitting results.
We conclude that such a large gap is unlikely to arise as a result of  possible magnon-phonon coupling, since the magnon-phonon coupling induced gap always occurs at the intersection between the magnon and phonon bands and it  does not extend over the entire BZ.
It is thus expected that the gap in CrGeTe$_3$ shares the same origin as in CrSiTe$_3$, namely the antisymmetric exchange interaction e.g. like the SOC-mediated DM interaction.
Hence, it appears natural that a larger gap opening in CrGeTe$_3$ must arise from the larger SOC in the latter material when Si is replaced by  heavier Ge.

In order to determine the nature of magnon bands and the underlying magnetic exchange interactions in CrXTe$_3$ systems, we carry out LSWT calculations based on a generalized Heisenberg spin Hamiltonian including DM interaction and single-ion magnetic anisotropy  \cite{DosSantos2018}:

\begin{eqnarray}
\label{Hamiltonian}
\begin{aligned}
H=& -\sum_{i<j} J_{ij} (\mathbf{S}_{i} \cdot \mathbf{S}_{j})-A_{zz}\sum_{i} (S_{i}^z)^2\\
&-\sum_{i<j} \mathbf{D}_{i j} \cdot\left(\mathbf{S}_{i} \times \mathbf{S}_{j}\right),
\end{aligned}
\end{eqnarray}

\noindent where $J_{ij}$ is the Heisenberg exchange constant between spins $\mathbf{S}_i$ and $\mathbf{S}_j$, $\mathbf{D}_{ij}$ represents the vector of the DM interaction between two magnetic ions when the inversion symmetry is broken, and $A_{zz}$ is the easy-axis anisotropy along the $c$ axis.
Note that we take $S=3/2$ given that the measured moment of Cr ions is extracted as $\sim$3 $\mu_B$ from the isotherm magnetization curve  shown in Fig.S1.
As illustrated in Fig.1(b), we  consider the 1st- and 2nd-NN Heisenberg exchange couplings in every honeycomb plane labeled as $J_{ab1}$ and $J_{ab2}$, and the 1st- and 2nd-NN exchange couplings between layers labeled as $J_{c1}$ and $J_{c2}$.
From the symmetry analysis, there will be no DM interaction between the nearest neighbors (NN), because of the space inversion.
However, the space inversion symmetry is broken between the 2nd NN, which can lead to a non-zero 2nd-NN DM interaction.


After a proper fitting process of the experimentally observed magnon spectra along the high-symmetry directions, the best-fit values for the exchange parameters of CrSiTe$_3$ and CrGeTe$_3$ were extracted.
As shown in Table.1, we find that the intra-layer interactions in CrXTe$_3$ are mainly dominated by the NN exchange couplings $J_{ab1}$ which are one order of magnitude larger than the interlayer exchange. Given the presence of sizeable interlayer exchange interactions, it is worth noting that CrXTe$_3$, strictly speaking, should be classified as quasi-2D spin systems.
While the direct exchange coupling between Cr ions is expected to be anti-ferromagnetic, due to the formation of Cr-Te-Cr bonds with a bond angle of nearly 90 degrees, the NN exchange couplings among Cr ions become ferromagnetic  ($J>0$) according to the Goodenough-Kanamori-Anderson rules~\cite{goodenough1963magnetism}.
Interestingly, although the fitting values of the inter-layer couplings $J_{c1}$ and $J_{c2}$ are quite comparable, the magnon dispersion along [00L] direction is likely much more sensitive to the 2nd-NN inter-layer coupling $J_{c2}$ because each Cr$^{3+}$ spin has nine 2nd-NNs but only one 1st NN.
Besides, the strength of the DM interactions is fitted as $|\mathbf{D}_{ij}|=0.12$ meV for CrSiTe$_3$ and $|\mathbf{D}_{ij}|=0.32$ meV for CrGeTe$_3$ which results in a $\sim$2 meV and $\sim$5 meV gap opening at the $K$ points, respectively.
We remark that the magnitude of the DM interaction in CrXTe$_3$ constitutes about 10\% of $J_{ab1}$ and it is quite comparable to the value of exchange  $J_{ab2}$  between 2nd-NN Cr ions.
From the isothermal magnetization curve at 2 K (Fig.S1), we can extract the magneto-crystalline anisotropy energy per Cr$^{3+}$, which is only 0.09 meV for CrSiTe$_3$ and 0.02 meV for CrGeTe$_3$, indicating a small single-ion anisotropy in the CrXTe$_3$ system.
In terms of the small single-ion anisotropy, as expected, we failed to determine the size of the gap at the $\mathit{\Gamma}$ point from our energy-scan data directly, and we used instead the quadratic fitting result 0.01 meV (shown in Fig.S5 and Fig.S6) for the anisotropy term $A$ in the Heisenberg-DM model.

As shown in Figs.2(d)-(f) and Figs.3(c)(d), the magnon bands of CrXTe$_3$ calculated from the Heisenberg-DM model are consistent with our experimentally observed spectra, taking into account a finite energy resolution.
Especially in the vicinity of the $K$ points, the opened gap is also well described by the model.
In addition, a range of constant-energy slices near the BZ boundary are found to be in an excellent agreement with the simulation results, as shown in Fig.S4.
Moreover, the calculated spectra also reproduce the characteristic weak/strong-intensity of the magnon bands in different BZs, except that the optical modes of CrGeTe$_3$ are significantly enhanced by the possible hybridization with the phonon modes, as discussed above.

The overall band dispersion is mainly determined by the ferromagnetic NN Heisenberg exchange couplings, and the gap opening at the $K$ points cannot be reproduced by simply adding exchange couplings with further neighbors in the model, until an anisotropic exchange coupling term such as the DM interaction is taken into consideration.
It is known that the DM interaction acts as an effective gauge potential for magnonic states and it can open topologically nontrivial band gaps in a magnonic system \cite{Owerre2016c,Kim2016}.
Without the  DM interaction, a crossing point located in the vicinity of $K$ point would be present,  Fig.4(a), while the gap
is opened immediately as the DM interaction is introduced, with the band gap size increasing almost linearly with the strength of the DM interaction,  Fig.4(b).
Overall, the contructed Heisenberg-DM model reproduces spin waves in CrXTe$_3$ very nicely, with one key point being that the DM interaction in our case is not only dependent on the SOC strength of Te atoms along the Cr-Te-Cr paths, but it is also influenced by  Si/Ge atoms.
The large difference of the gap size in CrXTe$_3$ thus translates into a possibility to tune the DM interaction by the substitution of the atoms with different SOC strength in the honeycomb center.

Interestingly, in addition to the Heisenberg-DM model, introducing the Kitaev interactions can also produce similar spin wave spectra and induce a gap opening at the $K$ points.
If we exclude X atoms, the rest of CrXTe$_3$ shares a common atomic structure with the well-studied Kitaev spin liquid candidate $\alpha$-RuCl$_3$ \cite{Banerjee2016a}, and three kinds of diamond-shaped planes, composed of the nearest-neighbor Cr-Te-Cr bonds, are nearly orthogonal to each other.
In terms of the structure symmetry, it is possible to have an anisotropic exchange interaction between nearest neighbors, for instance, an anisotropic Ising interaction in the local basis that consists of the normal vectors of three Cr-Te-Cr planes, as shown in Fig.S7.
The presence of the additional anisotropy can make it possible to construct the Heisenberg-Kitaev Hamiltonian.
Surprisingly, in order to match the magnetic excitations in CrXTe$_3$, we found that the Kitaev exchange needs to be one order of magnitude larger than the NN Heisenberg term, as shown in Table.S1.
In the latter case, the fitted Kitaev exchange parameters of CrGeTe$_3$ are almost two times those of CrSiTe$_3$, which is quite interesting, since CrSiTe$_3$ and CrGeTe$_3$ share a similar local environment including the bond distance for NN Cr atoms.
One possible explanation would be that the $sp3$ hybridized orbitals of Si/Ge atoms in the honeycomb center are strongly coupled with the $p$ orbitals of Te atoms and subsequently strongly influence the strength of the Cr-Te-Cr super-exchange interaction.
Similar orbital hybridization effect is not rare since the super-exchange interaction mediated by nonmagnetic ions is also proposed in inverse-trirutile compound \cite{Zhu2014} and double-perovskites \cite{Katukuri2020}.
In any case, if CrXTe$_3$ is  dominated by the Kitaev interaction, then a strong in-plane anisotropy should be easily evident in the magnetization measurements, which has never been reported.
We thus conclude that the Heisenberg-Kitaev model is not a proper model for CrXTe$_3$.
More details concerning the discussion of Heisenberg-Kitaev model are given in the supplement.

Having determined the microscopic spin Hamiltonians for CrXTe$_3$, we turn to the analysis of the topological nature of the magnonic bands, and reveal theoretically the presence of a large thermal Hall effect, an unambiguous experimental signature of topologically nontrivial in-gap edge states. Our numerical calculations confirm a nonzero Berry curvature at the $K$ points and also a nonzero Chern number, indicating nontrivial topology of these magnon bands.
For bulk CrXTe$_3$, the total Chern number is determined to be $-3$ and $+3$ for the acoustic and optical magnon bands respectively, which means there is one topological edge state at the boundary of each honeycomb layer.
It is important to note that the sign of the Chern number depends on the magnetization direction which in our case is the $z$-axis.
The sign of the Chern number can be changed by reversing the magnetization direction, which can be naturally achieved at the magnetic domain walls in CrXTe$_3$ acting as perfect waveguides for the topological edge states.
In addition to the bulk system, we also investigated the topological properties of the monolayer system, where we observe an almost identical BZ Berry curvature distribution and predict only one acoustic branch with Chern number $-1$ and one optical branch with Chern number $+1$ for the magnon bands.
Due to the lack of any interlayer interactions, the magnon bands in the monolayer are not split as in the bulk, but the topological properties are sustained as long as the intra-layer DM interaction exists.
To simplify the calculation, we use the monolayer system to study the edge states.
The edge states of zigzag and armchair nanoribbons with the projections of the bulk states are shown together in Fig.4(c) and (d), where one can observe that the edges on the two sides of a nanoribbon hold their own exclusive edge states but with opposite propagation direction.
The role of the DM interaction here is similar to that of strong SOC in graphene-like systems which turns them into a quantum spin Hall insulator \cite{Kane2005}, and so CrXTe$_3$ can be also considered as a bosonic version of the well-known topological insulator state.
Correspondingly, the presence of nontrivial edge states can  contribute to the transverse thermal Hall conductivity to yield the topological thermal Hall effect when a longitudinal temperature gradient and out-of-plane magnetic field are applied.
As shown in Fig.4(e), our numerical estimates indicate that the thermal Hall conductivity generated by the edge states can reach the order of 10$^{-4}$ W/Km, which is large enough to be observed in experiment. This motivates the future confirmation of topological thermal Hall effect in CrXTe$_3$ from thermal transport experiments.

In summary, we carried out a very comprehensive inelastic neutron scattering study of the spin-wave excitations on the ferromagnetic honeycomb magnets CrXTe$_3$.
Our data shows a clear gap opening in the magnon bands at the K points, with the gap size of about 2 meV and 5 meV for CrSiTe$_3$ and CrGeTe$_3$, respectively.
The magnon band dispersion in CrXTe$_3$ can be described very well by either the Heisenberg-DM model or Heisenberg-Kitaev model.
The gap opening can be ascribed to the asymmetric DM interaction with an out-of-plane DM vector or Kitaev interactions.
However, due to the very small magnetic anisotropy observed in CrXTe$_3$, the Heisenberg-DM model, in which the isotropic Heisenberg exchange dominates, is perceived as the proper model for these materials.
The magnon spectra of CrXTe$_3$ are successfully reproduced by using the LSWT calculations.
The gap size was found to be proportional to the strength of the DM interaction.
The strength of the DM interaction is also strongly dependent on the SOC of the nonmagnetic Si/Ge atoms that are located at the honeycomb center.
Furthermore, our numerical calculations predict the nontrivial topological nature of the magnon bands gap at the $K$ points, and also reveal the existence of topological edge states at the sample boundaries and domain walls. In addition, we also calculated the reference values for the expected thermal Hall conductivity that may arise from the topological magnon edge states, which can be measured in future thermal transport experiments.
Based on our experimental and theoretical results, we propose that CrSiTe$_3$ and CrGeTe$_3$ present an  ideal platform to realize topological magnon insulators, in which the nontrivial magnon gaps can be intrinsically tuned by varying the SOC of nonmagnetic ions.
In contrast to CrI$_3$, the substitution of Si/Ge in CrXTe$_3$ system using a wide range of nonmagnetic ions like C, Si, Ge, Sn, Pb, Ga \cite{Chabungbam2017,Zhuang2015,Khan2020,Marfoua2021,Yu2018a} can in principle change the balance of the exchange interactions and thus enable the manipulation of the topological magnon gap size, leading to novel magnetic properties rooting in complex magnonic topology.

\section{MATERIALS AND METHODS}
High-quality single crystals of CrXTe$_3$ (X=Si,Ge) were grown by the flux method at the sample preparation laboratory of JCNS-MLZ in Garching.
Starting materials of Cr, X and Te were mixed in an Ar-filled glove box at a molar ratio of Cr:X:Te=1:1:10.
The mixture was placed in an alumina crucible, which was then sealed in an evacuated quartz tube.
The tube was heated up to 930 $^{\circ}$C (X=Ge) or 1030$^{\circ}$C (X=Si) in 4 hours and sustained there for 10 hours.
Then the tube was slowly cooled down to 600 $^{\circ}$C with a cooling speed of 3 $^{\circ}$C/h followed by separating the crystals from the Te flux by centrifuging.
Shiny and sizable crystals with hexagonal natural edges were obtained.

Single-crystal X-ray diffraction (XRD) was performed at room temperature with an incident wavelength of 1.54 \AA\ (Cu-K$_\alpha$) on a Bruker D2-Phaser X-ray diffractometer.
By using a SQUID magnetometer (Quantum Design), the temperature and field dependence of the magnetization of CrSiTe$_3$ and CrGeTe$_3$ were measured along both the $a$ and $c$ axes respectively.
The magnetic susceptibility was measured from 2 K to 300 K with both zero-field-cooling (ZFC) and field-cooling (FC) conditions for some selected single crystals.
The isothermal magnetization ($M-H$) curves were also measured in a sweeping field from $-$50 to 50 kOe at 2 K.

The polarized neutron diffraction measurement on single-crystal CrGeTe$_3$ was carried out at the cold-neutron polarized spectrometer DNS at MLZ (with $\lambda_i$ = 4.2 \AA) at T = 4 K.
Two-dimensional Q-maps in the (H,K,0) and (H,H,L) planes of reciprocal space at three different polarization modes were measured at 4 K.
For the measurements of spin-wave excitations, we co-aligned 2 pieces of large single crystals with a total mass of 1.3 g for CrSiTe$_3$, and more than 100 pieces with a total mass of about 1.4 g for CrGeTe$_3$.
All the inelastic neutron scattering experiments were performed by using a range of thermal and cold neutron triple-axis spectrometers, including PUMA at MLZ, FLEXX \cite{Le2013} at HZB, IN12, IN8 and IN22 at ILL.
Both the (H,H,L) and (H,K,0) oriented samples were prepared for each of CrSiTe$_3$ and CrGeTe$_3$.
The low energy magnon dispersion along [0,0,L] was measured at the cold neutron triple-axis at IN12 (with a fixed $k_f$ = 1.7 \AA$^{-1}$ and 2.8 \AA$^{-1}$) and at FLEXX (with $k_f$ = 1.55 \AA$^{-1}$).
The magnon dispersions along [H,H,0] and [H,0,0] were measured at the thermal neutron triple-axis spectrometers IN8 and PUMA (both with a fixed $k_f$ = 2.662 \AA$^{-1}$).
All the magnon dispersion data was collected at the base temperature of T = 2 K.

\section{SUPPLEMENTARY MATERIALS}
\noindent Detailed magnetic properties of CrSiTe$_3$ and CrGeTe$_3$.\\
Detailed analysis of the polarized neutron scattering data for CrGeTe$_3$.\\
Detailed analysis of the inelastic neutron scattering data: constant-energy mapping for CrSiTe$_3$, and the determination of the magnon gap at the Brillouin zone center for CrXTe$_3$.\\
Detailed descriptions of the spin Hamiltonian for the Heisenberg-DM model and Heisenberg-Kitaev model.\\
The calculation of the magnon band topology for CrXTe$_3$.\\
Fig. S1. XRD and magnetic properties of CrSiTe$_3$ and CrGeTe$_3$. \\
Fig. S2. Polarized neutron scattering on CrGeTe$_3$. \\
Fig. S3. Mosaic width (i.e. FWHM) of the co-aligned CrXTe$_3$ samples. \\
Fig. S4. Constant-energy mappings of the magnon spectra of CrSiTe$_3$ in the (H,K,0) scattering plane measured at the thermal neutron triple-axis spectrometer PUMA. \\
Fig. S5. The magnon spectra of CrGeTe$_3$ measured at the cold neutron triple-axis spectrometer IN12.\\
Fig. S6. Cosine-function curve fittings of the respective magnon band dispersion of CrSiTe$_3$ and CrGeTe$_3$ along the [1,1,L] direction.\\
Fig. S7. Schematic plot for the Kitaev model.\\
Fig. S8. Comparison of the calculated magnon spectra with different models.\\
Fig. S9. The Berry curvature in the $K_x-K_y$ plane.\\
Fig. S10. Chern numbers of magnon bands.\\
Fig. S11. Calculated transverse thermal Hall conductivity.\\
References\cite{Alonso1995,Li2018b,Khan2019,Yuan2020,Casto2015,Kim2019,Zeisner2019,DosSantos2018,Gong2017,Holstein1940a,Singh2012,Banerjee2016a,Xu2018,Zhang2020c,Xu2020,Mook2014b,Onose2010a,Matsumoto2011a,Owerre2016b,Aguilera2020,Zhuang2015,Khan2020,Katsura2010}

\section{Acknowledgments}
\noindent \textbf{Acknowledgements:}
We acknowledge Susanne Mayr for assistance with the crystal orientation.
This work is based on the experiments performed at DNS, PUMA, IN8, IN12, IN22 and FLEXX neutron instruments.
\noindent \textbf{Funding:}
F.Z. acknowledges the funding support from the HGF–OCPC Postdoctoral Program.
Y.M. gratefully acknowledges the J\"ulich Supercomputing Centre and RWTH Aachen University for providing computational resources, as well as the support of Deutsche Forschungsgemeinschaft (DFG, German Research Foundation) $-$ TRR 173 $-$ 268565370 (project A11), TRR 288 – 422213477 (project B06).  
L.Z gratefully acknowledges the support of China Scholarship  Council  (CSC)  (Grant  No.  [2016]3100).
S. L. acknowledges the funding support provided by the Priority Programme SPP 2244 2D Materials Physics of van der Waals Heterostructures of the Deutsche Forschungsgemeinschaft (DFG) (project LO 1659/7-1) and the European Research Council (ERC) under the European Union’s Horizon 2020 research and innovation programme (ERC-consolidator Grant No. 681405 DYNASORE).
S.B. acknowledges financial support by the Deutsche Forschungsgemeinschaft (DFG) within CRC 1238 (project number 277146847, subproject C01).
\noindent \textbf{Author contributions: }
F.Z. and Y.S conceived the project.
Y.S., Y.M., S.B., T.B. co-supervised the work.
X.W. carried out the single-crystal growth.
F.Z, X.W., Y.S. carried out the neutron scattering experiments with supports from J.S., T.M., K.S., W.F.S., A.I., J.P., J.X., J.M..
L.Z. carried out the theoretical analysis with supports from F.J.D.S., S.L., Y.M..
F.Z. carried out the X-ray diffraction and magnetization measurements.
F.Z., L.Z., Y.S. wrote the manuscript with contributions from all authors.
\noindent \textbf{Competing interests: }
The authors declare that they have no competing interests.
\noindent \textbf{Data and materials availability: }
All data needed to evaluate the conclusions in the paper are present in the paper and/or the supplementary materials. The raw data taken at the Institut Laue-Langevin (ILL) can be accessed at https://doi.ill.fr/10.5291/ILL-DATA.4-01-1634 and https://doi.ill.fr/10.5291/ILL-DATA.CRG-2498. The raw data taken at the Heinz Maier-Leibnitz Zentrum (MLZ) and at the Berlin Neutron Scattering Center (BENSC) can be accessed at https://doi.org/10.5281/zenodo.4963138. 




\bibliography{CXT_ref}
%
\cleardoublepage

\begin{table}[b!]
\setlength{\tabcolsep}{12mm}{
\begin{tabular}{c@{~~~~}c@{~~~~}c@{~~~~}c@{~~~~}c@{~~~~}c@{~~~~}c}
    \toprule
       unit (meV) & $J_{ab1}$ & $J_{ab2}$ & $J_{c1}$ & $J_{c2}$ & $\Vec{D}_{ij}$ \\
    \midrule
        CrSiTe$_3$ & 1.49 & 0.15 & 0.07 & 0.06 & (0, 0, 0.12)  \\
    \midrule
        CrGeTe$_3$ & 2.73 & 0.33 & 0.10 & 0.08 & (0, 0, 0.32)  \\
    \bottomrule
\end{tabular}}
\caption{
\textbf{Spin Hamiltonian parameters}. The values of exchange interactions including the DM interactions are listed together for both CrSiTe$_3$ and CrGeTe$_3$.
The value of the 2nd-NN DM-vector was chosen to reproduce the experimental spin-wave dispersion.
The units of the parameters indicated here is meV.
The single-ion anisotropy is fixed to 0.01 meV.
}
\label{table:J}
\end{table}


\newpage

\begin{figure*}
\centering
\includegraphics[width=10cm]{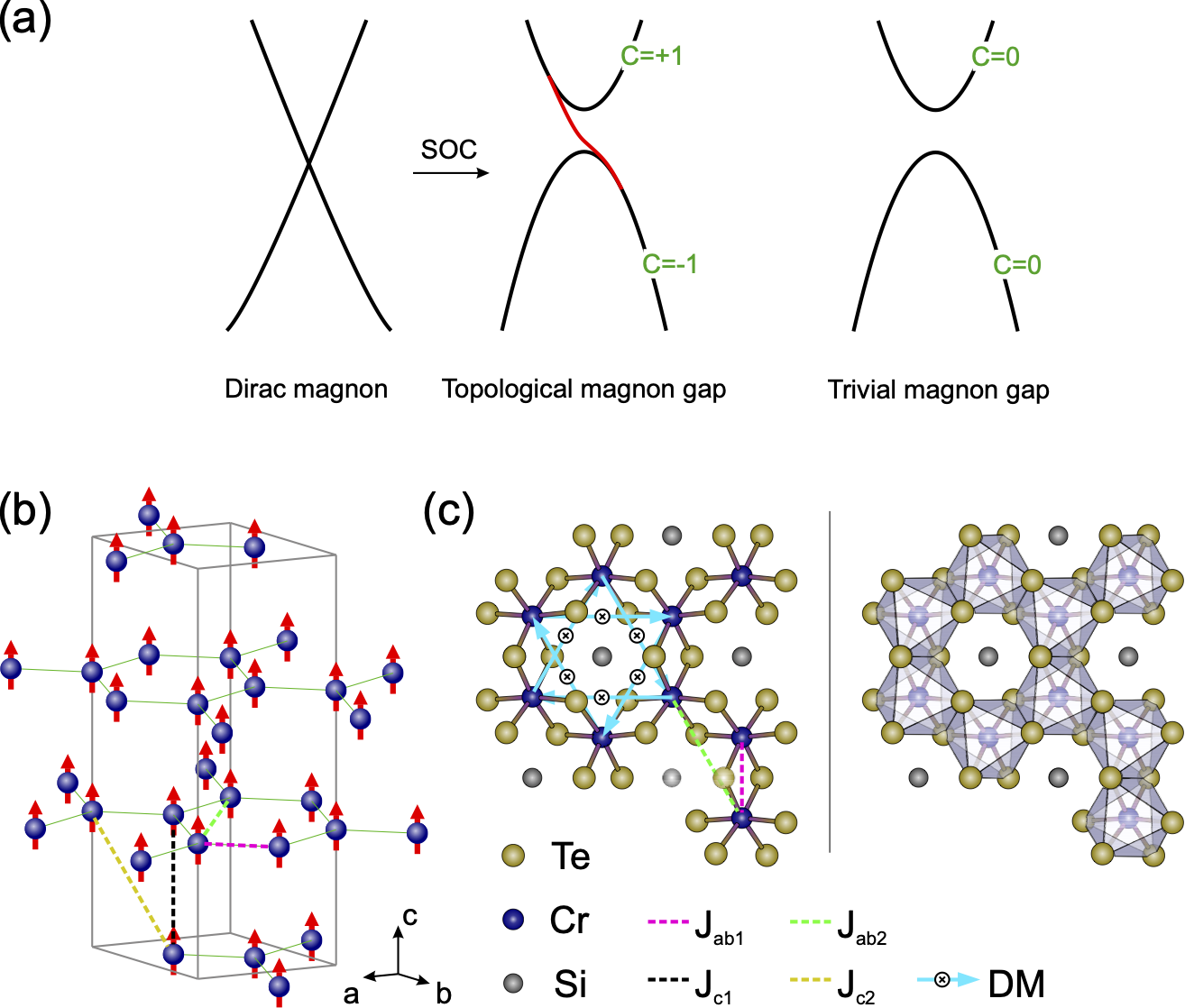}
\end{figure*}

\noindent {\bf Fig. 1.} 
\textbf{Schematic of the trivial and topological magnon bands, and of the magnetic and atomic structures of CrXTe$_3$}.
(a) Schematic of the band dispersion for a Dirac magnon, a topological magnon and a trivial magnon.
(b) Magnetic structure of CrSiTe$_3$.
In the $ab$-plane, the Cr atoms form a honeycomb lattice represented by the dark blue spheres and green solid lines.
The magnetic moments are represented by red arrows.
The first and second nearest-neighbor exchange interactions in intra- and inter-planes are represented by purple, green, black and yellow dash lines respectively.
(c) View perpendicular to the $ab$-plane showing the honeycomb network of CrSiTe$_3$.
The honeycomb network is caged by the edge-sharing octahedra composed of Te atoms and Si-Si dimers located in the center.
The light blue arrows represent the bond directions of the DM interactions between the second nearest-neighbor Cr atoms, and all the DM vectors share a common sign along the $c$ axis.

\newpage
\begin{figure*}
\centering
\includegraphics[width=16cm]{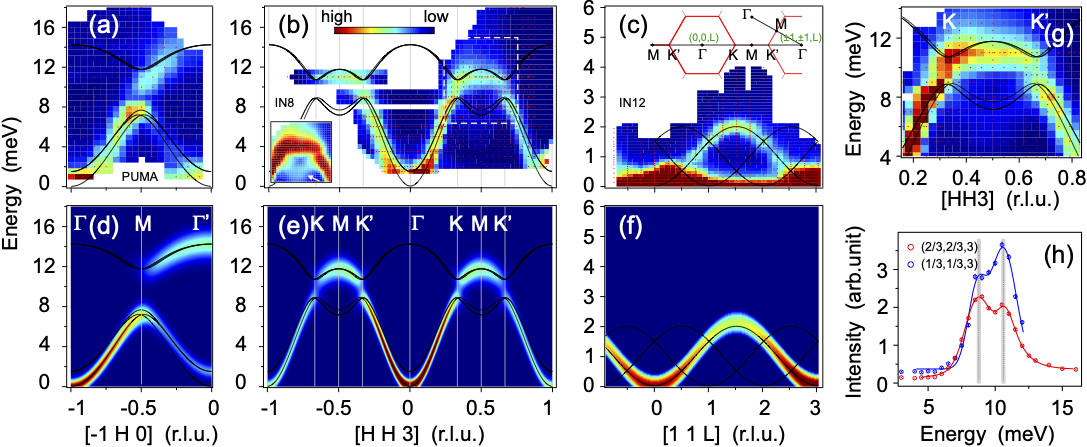}
\end{figure*}

\noindent {\bf Fig. 2.}
\textbf{Spin-wave excitations in CrSiTe$_3$}.
(a-c) Energy- and momentum-resolved neutron scattering intensity maps of magnons in CrSiTe$_3$ along the high-symmetry directions measured at the thermal neutron triple-axis spectrometer PUMA and IN8, and at the cold neutron triple-axis spectrometer IN12, respectively.
The black solid lines are the calculated magnon dispersion curves based on the parameters of Heisenberg-DM model presented in this paper.
The inset in (b) is a contrast adjusted plot for  the dashed rectangle part to make the acoustic branch easy to see.
The inset in (c) shows the exact scan paths in the reciprocal space.
(d-f) Calculated magnon spectra intensity maps for (a-c) respectively.
The calculated spectra are convolved with an energy resolution of 1 meV to compare with the experimental data.
(g) Zoom-in plot of the magnon spectra from (b) near the $K$ points.
(h) Energy scans of magnon density of states at the $K$ points.
The solid lines are the two-peak Gauss fitting results, and the fitted peak positions and error bars are indicated by the vertical dash lines with gray shadow.

\newpage
\begin{figure*}
\centering
\includegraphics[width=16cm]{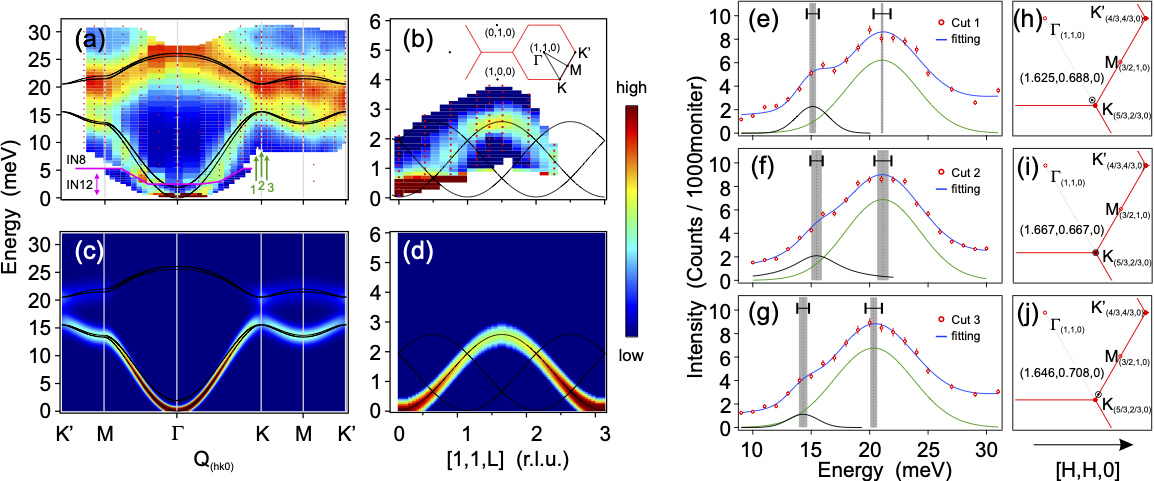}
\end{figure*}

\noindent {\bf Fig. 3.}
\textbf{Spin-wave excitations in CrGeTe$_3$}.
(a,b) Energy- and momentum-resolved neutron scattering intensity maps of magnon in CrGeTe$_3$ along the high-symmetry directions measured at IN8 and IN12, respectively.
Black solid lines are the calculated magnon dispersion curves.
Inset in (b) shows the projected BZ with high-symmetry points and the scan paths in the experiments.
(c,d) The corresponding calculated magnon spectra intensity maps for (a,b) by using the 2nd-NN DM interaction model.
The calculated spectra are convolved with an energy resolution of 1 meV to compare with experimental data.
(e-g) The line profiles of constant-Q energy scan at the positions of 1,2,3 marked by the green arrows in (a) near the $K$ point $Q$ = (5/3,2/3,0).
The solid lines are the multi-peak Gauss fitting results.
The peak positions and the errors are indicated respectively by the dash lines and the gray shadows, the corresponding energy resolutions are represented by the black horizontal bar with caps.
(h-j) The actual $Q$ positions of the Cut 1,2,3 are marked by black circles with center dots in the reciprocal space.
The red solid lines are the BZ boundaries.

\newpage
\begin{figure*}
\centering
\includegraphics[width=8cm]{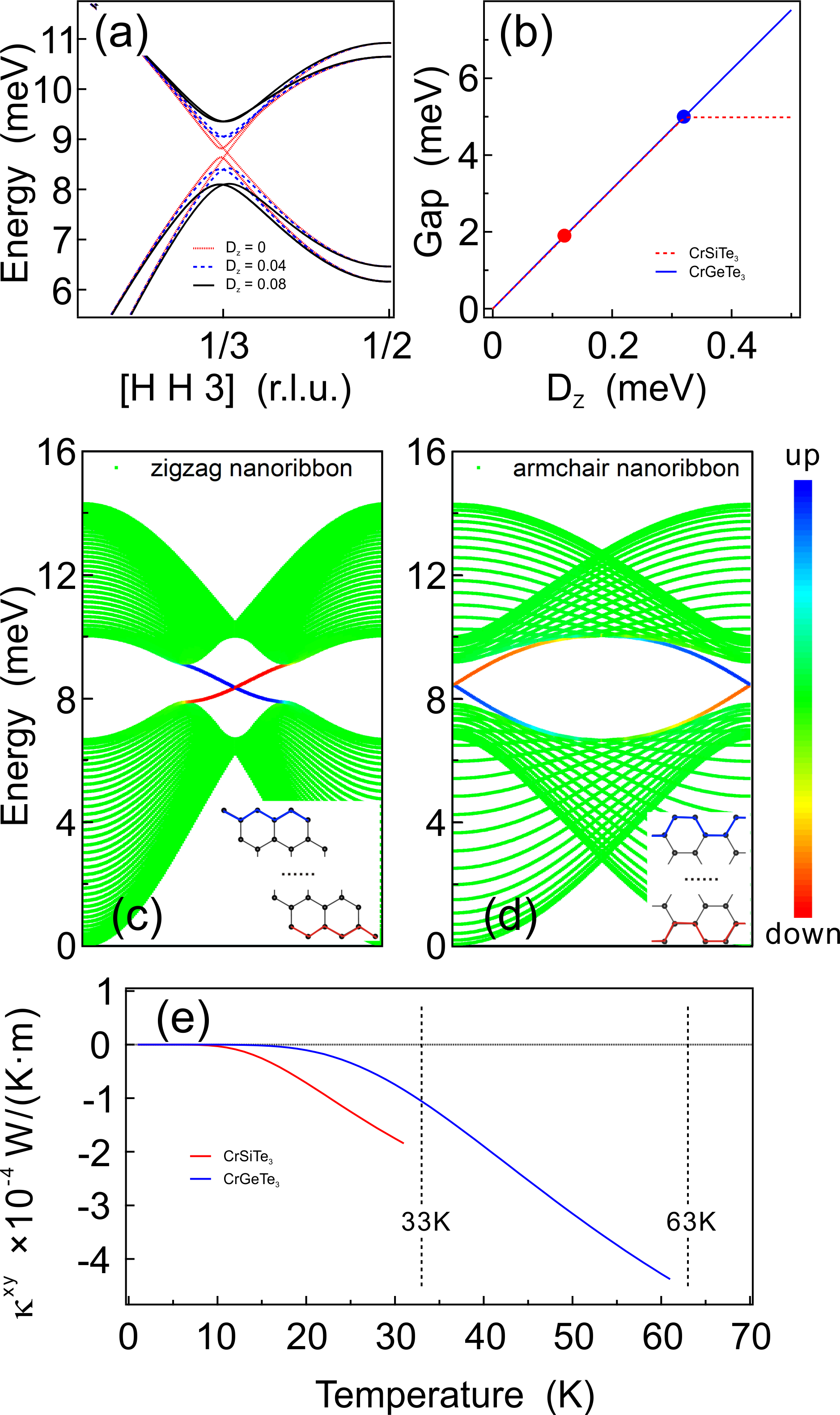}
\end{figure*}

\noindent {\bf Fig. 4.}
\textbf{The impact of the DM interaction on the magnon dispersion}.
The magnon dispersions of CrSiTe$_3$ with different DM interaction strength are compared in (a).
(b) The relationships between the opened global band gap and the strength of the DM interaction, the red and blue filled circles correspond to the values extracted from the magnon bands.
(c,d) The edge states of the monolayer CrSiTe$_3$ for the respective zigzag and armchair nanoribbon.
The color scale represents the weight of the magnonic wave function along the slab.
(e) Temperature dependence of the topological thermal Hall conductivity of CrSiTe$_3$ and of CrGeTe$_3$ in the ferromagnetic ordered phases.

\cleardoublepage

\begin{center}
\textbf{\large
Supplemental materials: Topological magnon insulators in two-dimensional van der Waals ferromagnets CrSiTe$_3$ and CrGeTe$_3$: towards intrinsic gap-tunability
}
\end{center}
\setcounter{equation}{0}
\setcounter{figure}{0}
\setcounter{table}{0}
\setcounter{page}{1}
\makeatletter
\renewcommand{\theequation}{S\arabic{equation}}
\renewcommand{\thefigure}{S\arabic{figure}}
\renewcommand{\thetable}{S\arabic{table}}

\begin{center}
Fengfeng Zhu, Lichuan Zhang, Xiao Wang, Flaviano Jos\'e dos Santos, Junda Song, Thomas Mueller, Karin Schmalzl, Wolfgang F. Schmidt, Alexandre Ivanov, Jitae T. Park, Jianhui Xu, Jie Ma, Samir Lounis, Stefan Bl\"ugel, Yuriy Mokrousov, Yixi Su, Thomas Br\"uckel
\end{center}

\section{DATA ANALYSIS}

\subsection{X-ray diffraction, magnetic properties and neutron scattering}

\begin{figure*}
\centering
\includegraphics[width=15cm]{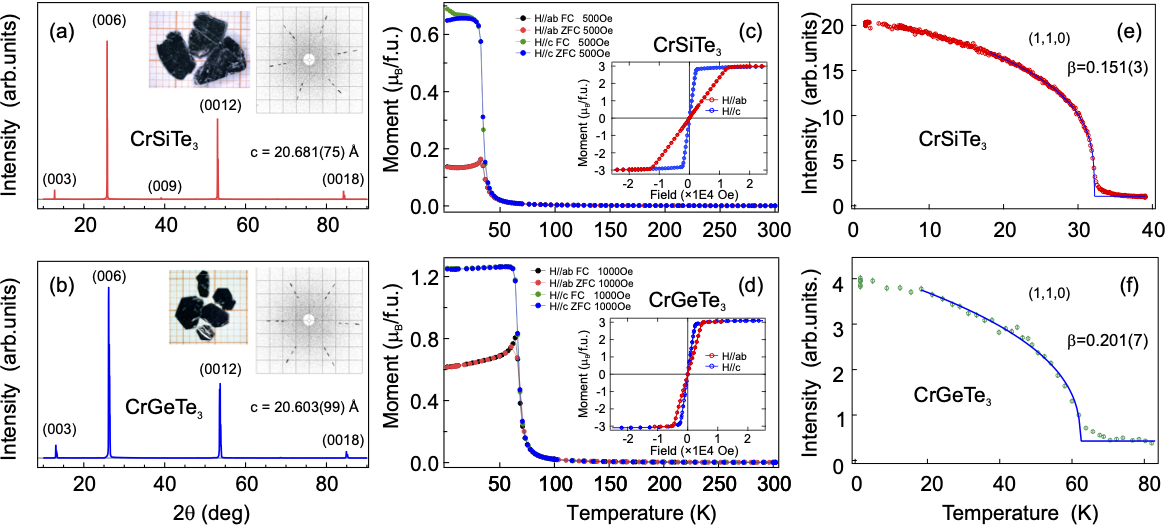}
\caption{\label{fig:SS_XRD_MTMH}
\textbf{X-ray diffraction and  magnetic properties of CrXTe$_3$}.
(a,b) X-ray diffraction of single-crystal CrSiTe$_3$ and CrGeTe$_3$ at 300 K.
The insets are the sample pictures and the corresponding X-ray Laue pattern of the (H,K,0) reciprocal plane.
(c,d) ZFC/FC magnetization curves measured under applied magnetic fields along the $c$ and $a$ axes.
Insets show the corresponding magnetization as a function of field at 2 K for $H\parallel c$ and $H\parallel ab$.
(e,f) Temperature dependence of the (1,1,0) magnetic Bragg peak intensity of CrSiTe$_3$ and CrGeTe$_3$ measured at DNS, MLZ.
The solid lines are the fittings of the experimental data in the vicinity of the ferromagnetic transition.
}
\end{figure*}

The lattice parameter $c$ = 20.681(75) \AA\ for CrSiTe$_3$ and $c$ = 20.603(99) \AA\ for CrGeTe$_3$ at room temperature extracted from XRD (in Fig.\ref{fig:SS_XRD_MTMH}[(a),(b)]) is quite consistent with the previous results \cite{Alonso1995,Li2018b,Khan2019}.
From the temperature dependence of the magnetic susceptibility (shown in Fig.\ref{fig:SS_XRD_MTMH}[(c),(d)]), the ferromagnetic transitions can clearly be observed at $T_c\approx$ 33 K for CrSiTe$_3$ and $T_c\approx$ 63 K for CrGeTe$_3$, respectively.
In the insets of Fig.\ref{fig:SS_XRD_MTMH}[(c),(d)], the isothermal magnetization ($M-H$) curves show small magnetic anisotropies, and the magneto-crystalline anisotropy is extracted as 0.09 meV for CrSiTe$_3$ and 0.02 meV for CrGeTe$_3$, respectively.
In Fig.\ref{fig:SS_XRD_MTMH}[(e),(f)], the phase transition is also observed from the temperature dependence of the (1,1,0) magnetic Bragg peak for both CrSiTe$_3$ and CrGeTe$_3$.
The temperature dependence of the order parameter were fitted with the power law equation $I = I_0 + A (1- \frac{T}{T_N})^{2\beta}$, the fitted transition temperatures are about 33 K for CrSiTe$_3$ and 63 K for CrGeTe$_3$, which are in a good agreement with our magnetization results.
The critical exponents are extracted as $\beta$ = 0.151(3) for CrSiTe$_3$ and $\beta$ = 0.201(7) for CrGeTe$_3$.

\begin{figure*}
\centering
\includegraphics[width=15cm]{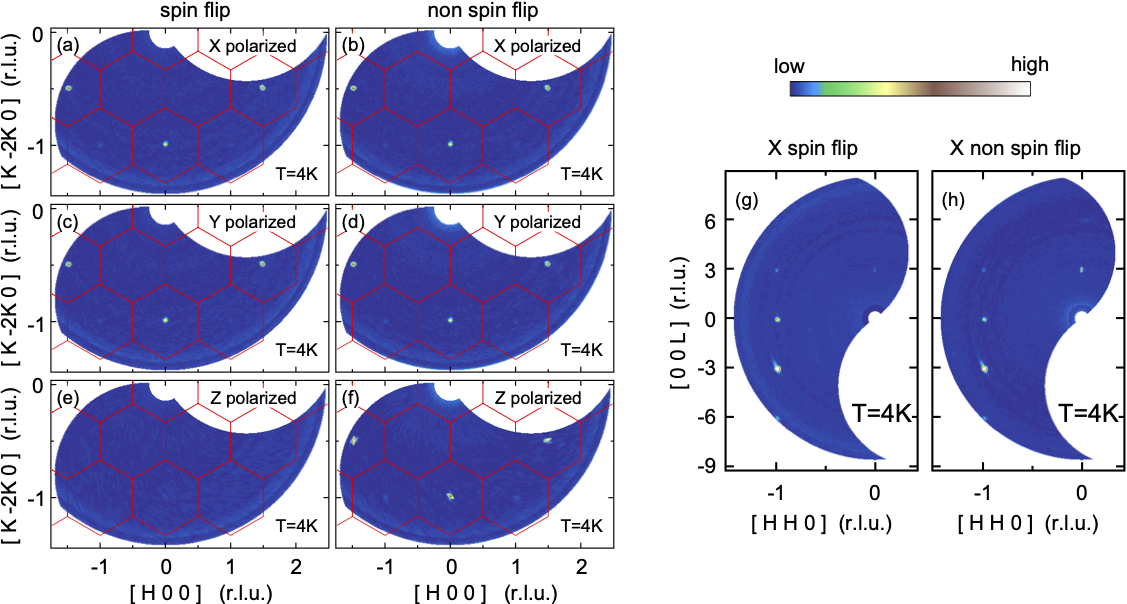}
\caption{\label{fig:SS_CGT_DNSPlot}
\textbf{Polarized elastic neutron scattering maps}. (a-f) $Q$ Maps in the (H,K,0) scattering plane and (g,h) in the (H,H,L) scattering plane measured on CrGeTe$_3$ at DNS.
Red solid lines are the Brillouin zone boundaries.
}
\end{figure*}

With polarized neutron diffraction, we can easily determine the directions of the ordered magnetic moments, because the sign of polarized neutrons can only be flipped when there exists a non-zero component of magnetic moments perpendicular to the polarization of the neutron beam $\textbf{P}$ and the scattering wave vector $\textbf{Q}$.
Due to a favorable situation related to the ferromagnetic domains in this sample, no neutron depolarization was noticed for the polarized measurements in the ferromagnetic state.
As shown in Fig.\ref{fig:SS_CGT_DNSPlot}, 2D Q-maps in the (H,K,0) and (H,H,L) planes of reciprocal space at three different polarization modes were measured at 4 K.
The diffraction patterns of $x$ and $y$ polarization for the respective spin-flip and non-spin-flip channels (Fig.\ref{fig:SS_CGT_DNSPlot}(a-d)) are basically the same in the (H,K,0) scattering plane, and all the nuclear and magnetic diffraction peaks relocated on the same position in the center of Brillouin zone, which indicates its ferromagnetic character and that the ferromagnetic moment has a non-zero component along the $z$ directions at least.
For the $z$ polarization, there are no diffraction peaks observed in the spin-flip channel but the same diffraction pattern as in the previous $xy$ polarizations of the non-spin-flip channel, as shown in Fig.\ref{fig:SS_CGT_DNSPlot}[(e),(f)], which confirms that the ordered ferromagnetic moments are aligned exclusively along the $c$ axis.
The absence of the (0,0,3$n$) magnetic reflections in the $x$ spin-flip channel compared to the $x$ non-spin-flip channel (shown in Fig.\ref{fig:SS_CGT_DNSPlot}(g)), also indicates the magnetic moments are aligned along the $c$ axis.

\begin{figure*}
\centering
\includegraphics[width=14cm]{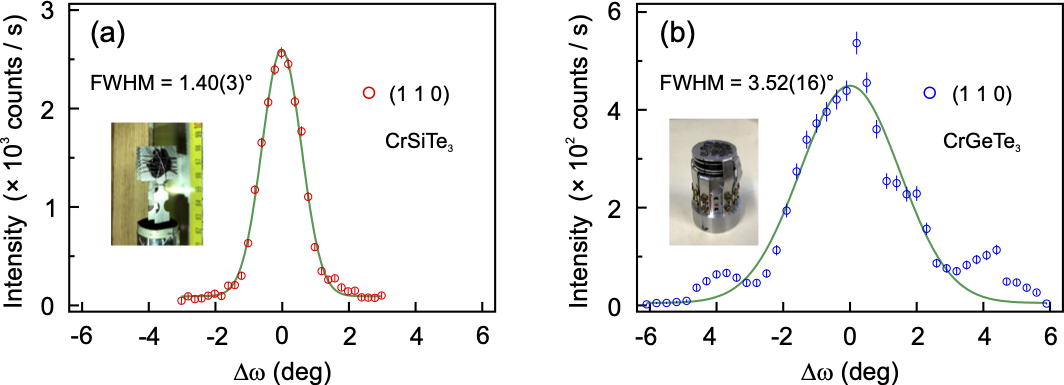}
\caption{\label{fig:SS_FWHM}
\textbf{Mosaic width of the co-aligned CrXTe$_3$ samples}.
(a),(b) Rocking curve scans of the (1,1,0) nuclear reflection at about 80 K show the alignment quality of CrSiTe$_3$ (with 2 pieces) and CrGeTe$_3$ (with more than 100 pieces) respectively.
The insets are the corresponding pictures of the samples measured at IN8.
}
\end{figure*}

For the measurements of spin-wave excitations, the full width at half maximum of CrSiTe$_3$ is much smaller than that of CrGeTe$_3$.
For CrSiTe$_3$ single crystals, due to the large size and thickness, we used only 2 single crystals and the FWHM of the co-aligned sample turns out to be only 1.4$^\circ$ in the rocking curve scan at IN8, as shown in Fig.\ref{fig:SS_FWHM}(a).
However, for CrGeTe$_3$ single crystals, their usual dimension is about 2 mm $\times$ 2 mm $\times$ 0.5 mm, so we co-aligned about 0.8 g samples on 5 horizontal aluminum plates and about 0.6 g samples on 3 vertical aluminum plates for the experiments with (H,K,0) and (H,H,L) as scattering planes, respectively.
The corresponding FWHM of the (H,K,0) orientated CrGeTe$_3$ sample is about 3.52$^\circ$ (shown in Fig.\ref{fig:SS_FWHM}(b)).

\subsection{Constant-energy mapping around the $K$ points}
\begin{figure*}
\centering
\includegraphics[width=15cm]{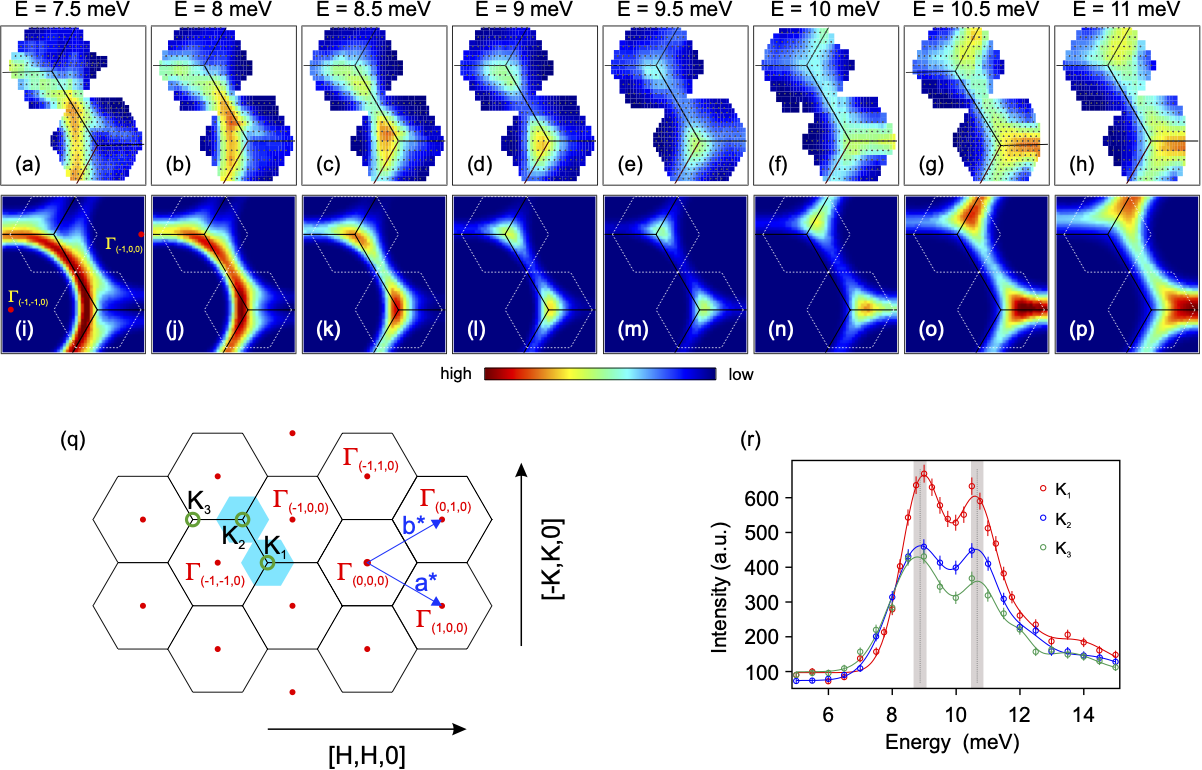}
\caption{\label{fig:CST_PUMA_CE}
\textbf{Spin-wave excitations of CrSiTe$_3$ in the (H,K,0) scattering plane}.
(a-h) Constant-energy mappings of the magnon spectra of CrSiTe$_3$ in the (H,K,0) scattering plane measured at the thermal neutron triple-axis spectrometer PUMA.
The energies of the constant-energy mappings are chosen to cover the whole energy scale of the opened gaps at the $K$ points with an energy step of 0.5 meV.
The black solid lines denote the boundaries of the 2D BZs.
The dark gray dots represent the actually measured $Q$ points.
(i-p) Calculated constant-energy mappings using the parameters of the 2nd-NN DM interaction model in this paper.
The calculated spectra are convolved with the estimated instrument energy resolution of 1.5 meV.
Areas inside the white dashed lines corresponds to the $Q$ range for the experimental data.
(q) Schematics of the projected 2D BZs.
The BZ center $\mathit{\Gamma}$ points and 3 selected $K$ points are denoted by the red dots and green circles, respectively.
The light blue hexagon is a schematic for the corresponding $Q$ positions for the experimental data in (a-h).
(r) Constant-Q energy scans of the magnon spectra at 3 different $K$ points in (q).
The solid lines are the multi-peak Gauss fitting.
The dash lines and grey shadow represent the averaged energy positions and error bars for the magnon bands at the $K$ points.
}
\end{figure*}

To gain more information about the the magnon band dispersion in the vicinity of the $K$ points, we have also very carefully measured the constant-energy scan contour maps in the (H,K,0) reciprocal plane with various energy transfers, as shown in Fig.\ref{fig:CST_PUMA_CE}(a-h).
The energy transfers were selected from 7.5 meV to 11 meV with a step of 0.5 meV to cover the whole range of the gap opening at the $K$ points.
In Fig.\ref{fig:CST_PUMA_CE}(q), a 2D reciprocal space map with BZ boundaries and high-symmetry points in (H,K,0) is shown, and the measured regions of the constant-energy slices in $Q$ space are indicated by the light blue hexagons.
However, the spectral function (or dynamic structure factor) is strongly dependent on the momentum transfer, causing intensity enhancement or suppression in some specific parts of the contour maps.
According to the previously calculated spectral function for the honeycomb ferromagnet with DM interactions, the magnon shows a shorter lifetime in the acoustic mode on one side of the $K$ point and, in the optical mode on the opposite side of $K$, which is consistent with the antisymmetric intensity distribution that we observed in Fig.\ref{fig:CST_PUMA_CE}(a-h).
From the constant-energy slices, the 2D projected magnon ``cone'' is observed to become weaker and smaller towards the gap energy, and become larger again by increasing energy transfers.
In betweens, for example at $E$ = 9.5 meV, only discrete spots with weak intensity can be seen at the $K$ points, as shown in Fig.\ref{fig:CST_PUMA_CE}(e).

Although the existence of a gap opening is almost clear by observing the intensity change of the constant-energy contours, it is not rigorous and convincing enough, given that similar constant-energy contours were also observed in the topological Dirac magnon material CoTiO$_3$ where there exists no gap at the $K$ points \cite{Yuan2020}.
To exclude the involvement of the Dirac magnon, further detailed constant-Q energy scans at various $K$ points are quite necessary.
In Fig.\ref{fig:CST_PUMA_CE}(r), the energy-scan line profiles were extracted at three $K$ points [highlighted by green circles in Fig.\ref{fig:CST_PUMA_CE}(q)] with different momentum transfers, and all of them show two peaks located at around 8.7 meV and 10.7 meV, which clearly indicate a 2 meV gap opening at around 9.5 meV.
Due to the limited energy resolution of $\sim$ 1.3 meV that is also comparable to the gap size, the in-gap intensity is not exactly zero in our experimental configuration.
By applying a convolution of the instrument energy resolution, similar constant-energy patterns including the discrete spot-shaped contours at $E$ = 9.5 meV can be very well reproduced in our simulations, as shown in Fig.\ref{fig:CST_PUMA_CE}(i-p).

\subsection{Determination of the magnon gap at the Brillouin zone center}

\begin{figure*}
\centering
\includegraphics[width=14cm]{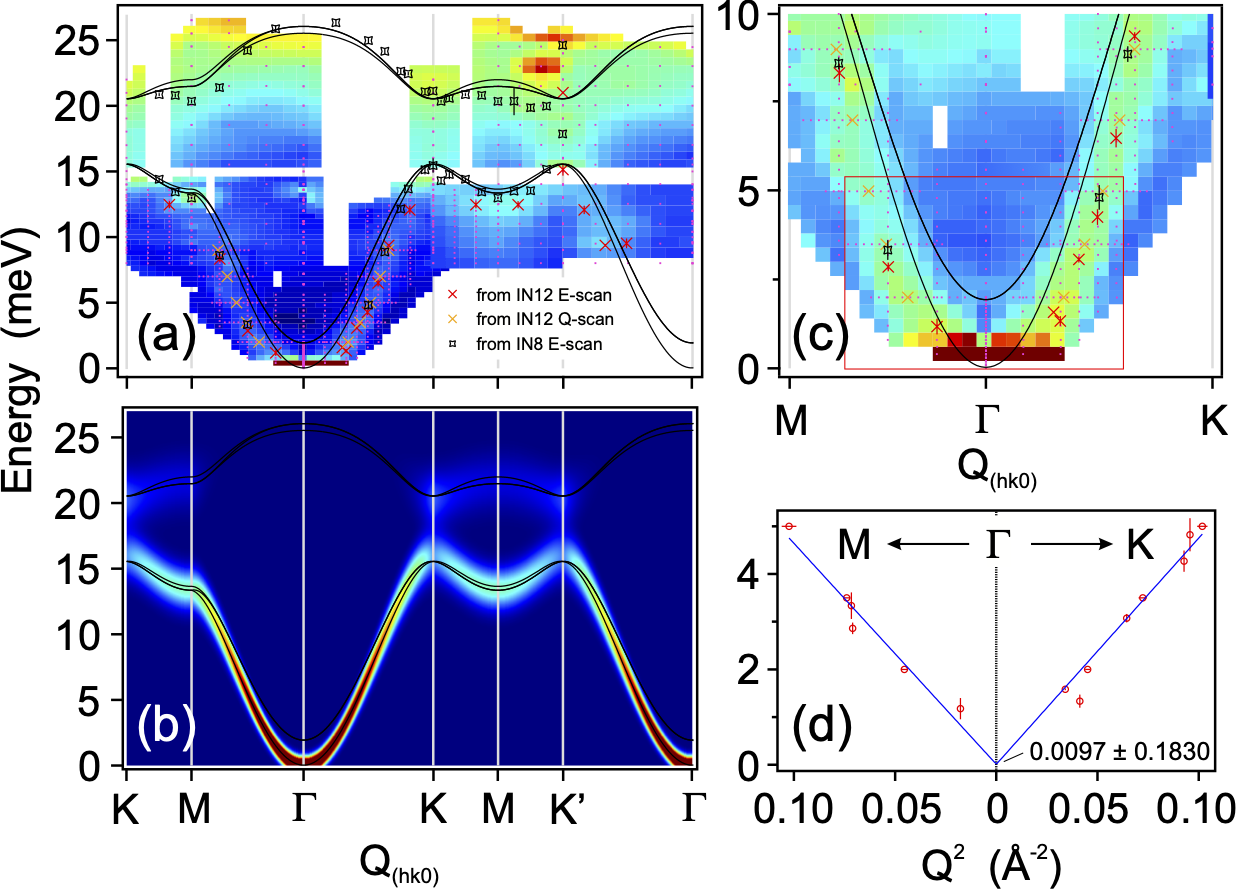}
\caption{\label{fig:CGT_IN12}
\textbf{The low-energy magnon branches of CrGeTe$_3$}.
(a) The magnon spectra of CrGeTe$_3$ measured at the cold neutron triple-axis spectrometer IN12 (with a fixed $k_f$ = 1.7 \AA$^{-1}$ and 2.8 \AA$^{-1}$).
The solid lines are the calculated magnon dispersion curves.
The isolated crosses and stars are the fitted peak positions of various constant-Q and constant-E scans.
(b) The corresponding caculated magnon spectra by using the 2nd-NN DM interaction model.
(c) Enlarged plot of the low energy excitations in (a).
(d) Quadratic fitting of the magnon band for the low energy parts.
Only the peak positions inside the red rectangle in (c) are included in the fitting.
}
\end{figure*}

We tried to measure the spin wave gap at the $\mathit{\Gamma}$ point by using one of the best cold neutron triple-axis spectrometers, IN12, but we failed although the energy resolution and background level were quite good.
Unlike CrI$_3$, the anisotropy in both CrSiTe$_3$ and CrGeTe$_3$ is very small \cite{Casto2015,Kim2019,Zeisner2019}, it is really hard to resolve it directly from our low energy scan data.
To give out a reference value of the spin wave gap, we carefully measured the low energy magnon bands along the high-symmetry directions at 2 K.
In the color map of Fig.\ref{fig:CGT_IN12}(a), the magnon bands of CrGeTe$_3$ below 15 meV were measured at IN12 with $k_f$ = 1.7 \AA$^{-1}$, and the magnon bands above 15 meV were measured with $k_f$ = 2.8 \AA$^{-1}$.
Fig.\ref{fig:CGT_IN12}(b) shows the corresponding calculated dispersion curves of the same $Q$ path as in Fig.\ref{fig:CGT_IN12}(a) by using the fitted exchange parameters in the Heisenberg-DM model.
In the vicinity of the $\mathit{\Gamma}$ point at low energy, the magnon bands dispersions along $\mathit{\Gamma}-M$ and $\mathit{\Gamma}-K$ can be both fitted by the parabolic equation under the long-wave approximation.
Here, for the quadratic fitting, the Gauss fitted peak postions of the measured magnon bands below 5 meV were included, as shown in the red rectangular frame of Fig.\ref{fig:CGT_IN12}(c).
The fitting results were shown in the form of the energy as a linear function of the squared moment transfer $Q$, and the fitted gap size at the $\mathit{\Gamma}$ point is about 0.0097 meV but with a very large error of 0.183 meV, which indicates the gap of CrGeTe$_3$ is rather small and really hard to resolve by triple-axis spectroscopy.

\begin{figure*}
\centering
\includegraphics[width=13cm]{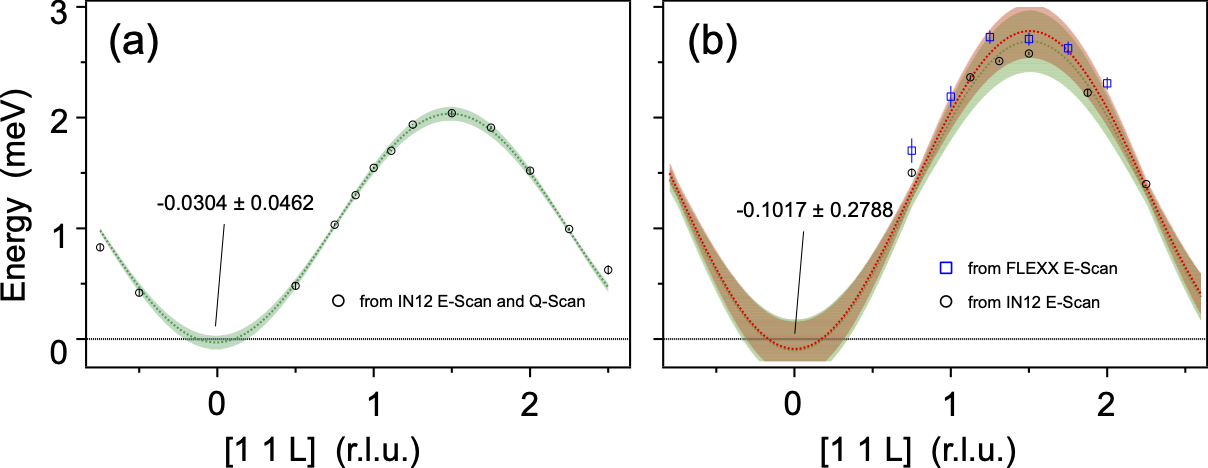}
\caption{\label{fig:SS_G_Gap}
\textbf{Fitting of the magnon dispersion along the [1,1,L] direction}.
(a),(b) Cosine-function curve fittings of the respective magnon band dispersion of CrSiTe$_3$ and CrGeTe$_3$ along the [1,1,L] direction.
The green dashed lines in (a,b) are the fittings according to the data collected from IN12 ($k_f$ = 1.7 \AA$^{-1}$).
The red dashed line in (b) is a combined fittings with the restriction in Fig.\ref{fig:CGT_IN12}(d) according to all the data of CrGeTe$_3$ collected from IN12 and FLEXX.
The light shading zones represent the confidence interval with a width of 2 standard errors.
}
\end{figure*}

For the same reason, the magnon band dispersions of CrSiTe$_3$ and CrGeTe$_3$ along the [0 0 L] direction were also fitted to gain a reference value of the spin wave gap at the $\mathit{\Gamma}$ point, as shown in Fig.\ref{fig:SS_G_Gap}.
Given the magnetic structure is ferromagnetic in both the intralayers and also the vdW interlayers, we can treat every three ABC-stacked ferromagnetic Cr$^{3+}$ as a single imaginary spin with $S=3/2\times3$, so the spin wave dispersion along the $c$ axis that is concerned here can be approximately described by using a one-dimensional ferromagnetic model with an effective nearest-neighborer ferromagnetic exchange interaction $\tilde{J}$:

\begin{eqnarray}
\label{Hamiltonian_1D}
\begin{aligned}
H=& -\tilde{J}\sum_{i}  (S_{i}^{x}S_{i+1}^{x}+S_{i}^{y}S_{i+1}^{y}+\Delta S_{i}^{z}S_{i+1}^{z})-A_{zz}\sum_{i} (S_{i}^z)^2,\\
\end{aligned}
\end{eqnarray}

\noindent where $\Delta$ and $A_{zz}$ denote the exchange and single-ion anisotropy, respectively.
In the case $\Delta = 1$, the exchange interaction reduces to the isotropic Heisenberg interaction.
At low temperatures ($k_BT\ll \tilde{J}$), the perturbations of the ground state are very small, so we can ignore the high-order terms and just keep the ground-state energy term and quadratic terms in bosonic operators.
Finally, the ferromagnetic magnon dispersion can be wrote as

\begin{eqnarray}
\label{Magnon_1D}
\begin{aligned}
\epsilon_k=& 2\tilde{J}S(\Delta-\cos(kc))+2A_{zz}S,\\
\end{aligned}
\end{eqnarray}

\noindent where $c$ is the lattice constant along the $z$ direction.
The gap at the Brillouin zone center $\mathit{\Gamma}$ point is expressed as $2\tilde{J}S(\Delta-1)+2A_{zz}S$.
The band dispersions in Fig.\ref{fig:SS_G_Gap} were fitted by using Eq.\ref{Magnon_1D}, and the effective exchange interactions $\tilde{J}$ were determined as 0.11(1) and 0.16(2) meV for CrSiTe$_3$ and CrGeTe$_3$ respectively.
From the fitting results, we failed to get the anisotropy gap size because of the extremely small magnetic anisotropy in CrXTe$_3$.
To give an idea of how small the single-ion anisotropy is, we set $\Delta =1$, namely Heisenberg interaction, and the single-ion anisotropy is estimated as $A_{zz}<0.0018$ meV for CrSiTe$_3$ and $A_{zz}<0.020$ meV for CrGeTe$_3$.
The small single-ion anisotropy can basically be neglected, compared to the large in-plane exchange interactions $J_{ab}$.
However, for the overall magnon band fitting (e.g. in Fig.\ref{fig:CGT_IN12}(b)), we symbolically set the single-ion anisotropy $A$ as 0.01 meV for the Hamiltonian in Eq.\ref{Hamiltonian_1D}.

\section{SPIN HARMILTONIAN}

\subsection{Heisenberg-DM model}
For the CrXTe$_3$, although the DM interaction will cancel because of the space inversion between the nearest-neighbor Cr atoms, the second nearest-neighbor interaction, the space inversion symmetry is actually broken in a honeycomb lattice, leading to a non-zero DM interaction.
Although the Te atoms may appear to have the strongest SOC in CrXTe$_3$, the magnitude of the DM interaction is more likely determined by the competition of one Cr atom and two Si/Ge atoms on two sides of the second nearest-neighbor Cr-Cr bond.
Here we use LSWT to calculate the magnon spectra of CrXTe$_3$, starting from the generalized Heisenberg model \cite{DosSantos2018},

\begin{equation}
  H=\sum_{i<j}\mathbf{S}_i^{\dagger}\hat{J}_{ij}\mathbf{S}^{\phantom{\dagger}}_j
  \label{equ1}\vspace{-1ex},
\end{equation}

\noindent where the interaction tensor between the lattice sites $i$ and $j$

\begin{equation}
\hat{J}_{ij}={
\left(
 \begin{array}{ccc}
 J_{ij}^{x}  &  D_{ij}^{z} &  -D_{ij}^{y}\\
 -D_{ij}^{z} &  J_{ij}^{y}  & D_{ij}^{x}\\
 D_{ij}^{y}  & -D_{ij}^{x}  & J_{ij}^{z}
 \end{array}
  \right)}
\end{equation}

\noindent includes the symmetric exchange $\mathbf{J}_{ij}$ and the antisymmetric off-diagonal DM interaction terms $\mathbf{D}_{ij}$, caused by the spin-orbit coupling.
The DM interaction vector is defined as $\mathbf{D}_{ij}=(D_{ij}^x, D_{ij}^y, D_{ij}^z)$.
Based on the symmetry and the Moriya's rule, the direction of the DM interaction vector is determined as out of plane.
Similar to earlier works \cite{Gong2017}, we include five Heisenberg exchange interactions in total, three in the intra-layers and two between the inter-layers, with their numerical values listed in Table~\ref{table:J}.

According to the LSWT, the Holstein-Primakoff transformation \cite{Holstein1940a} is adopted for the quantum spin operators.
Followed by the Fourier transformation for the boson operators, the Hamiltonian matrix in momentum space is obtained.
Then the eigenvalue and eigenvector, namely the magnon band dispersion can be extracted through diagonalizing the matrix.
As we expected, the band opens a gap at the $K$ point which is consistent with our experimental results.

\subsection{Heisenberg-Kitaev model}

\begin{figure*}
\centering
\includegraphics[width=14cm]{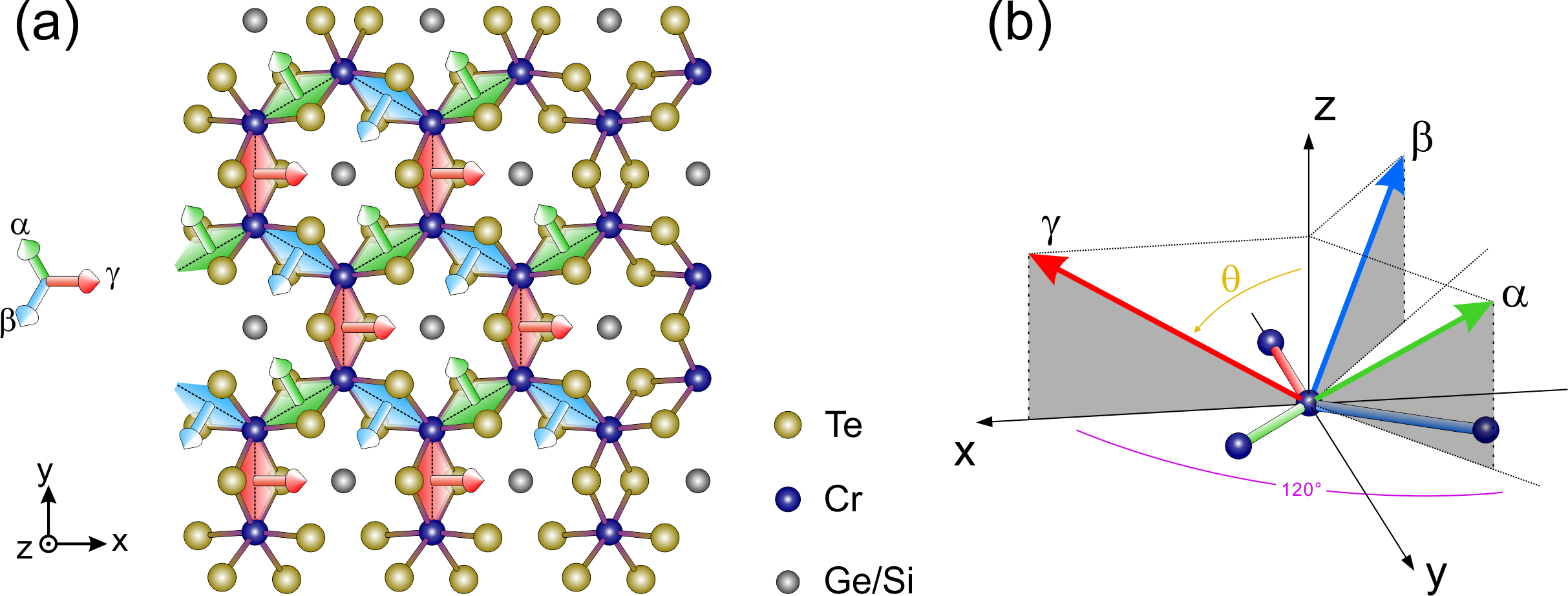}
\caption{\label{fig:SS_Models}
\textbf{The proposed Kitaev model for CrXTe$_3$}.
(a) Schematic plot for the Kitaev model.
The red, blue and green rhombus planes marked with perpendicular arrows are almost orthogonal to each other, which are used to represent 3 different bonds of the Kitaev model.
(b) The configuration of the local $\left \{ \alpha\beta\gamma \right \} $ coordinate in the global $\left \{ xyz \right \} $ coordinate.
}
\end{figure*}

\begin{figure*}
\centering
\includegraphics[width=12cm]{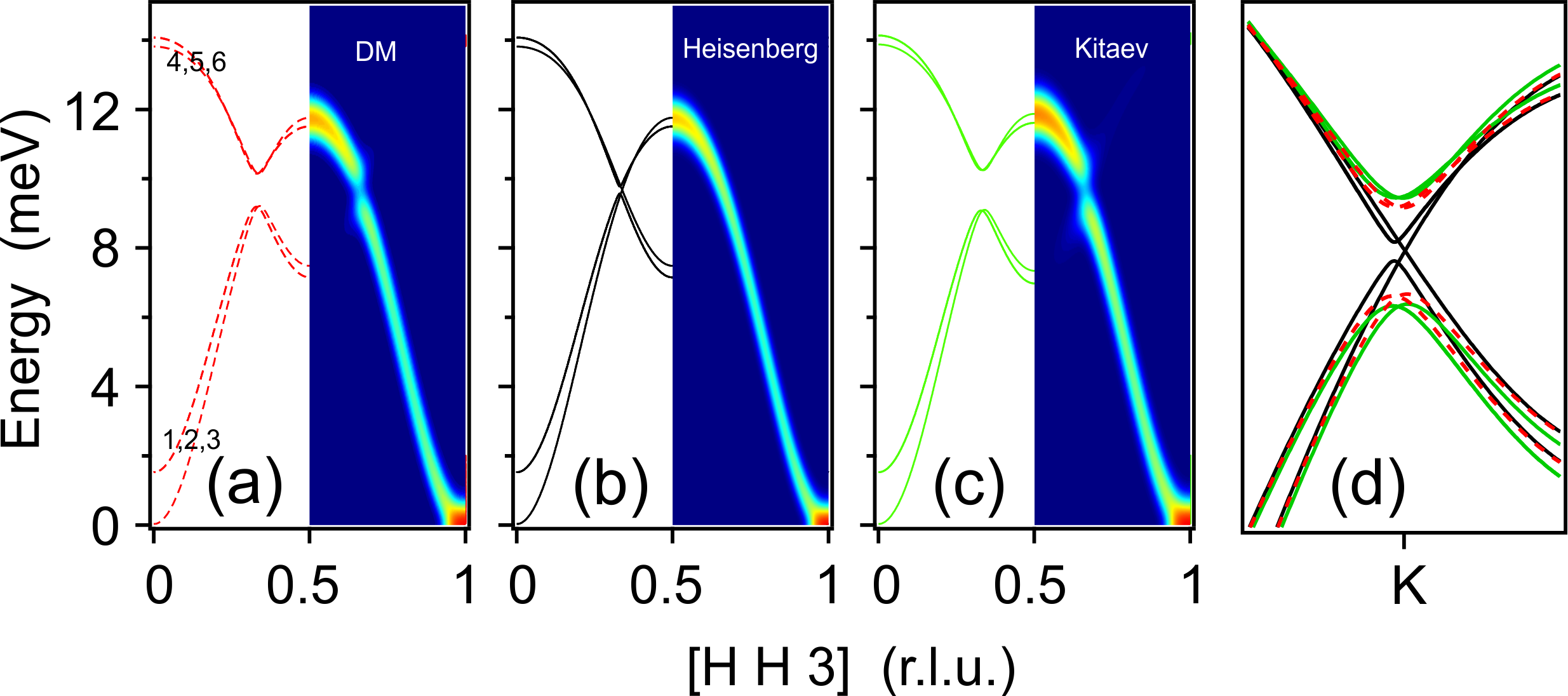}
\caption{\label{fig:SS_Models_Band}
\textbf{Comparison of the calculated magnon spectra with different models}.
(a) Magnon spectra calculated by using the Heisenberg with 2nd-NN DM interaction model.
Dash lines are the calculated magnon spectra at the BZ boundaries, and the intensity maps are the convoluted results with energy resolution of 1 meV.
(b) Magnon spectra calculated by using a simple Heisenberg model.
(c) Magnon spectra calculated by using the Heisenberg with a modified Kitaev interaction model.
(d) Comparison of the magnon band dispersion near the Dirac point between all the models in (a-c).
}
\end{figure*}

The Kitaev interaction is introduced to understand the magnetic behaviors that are close to quantum spin liquids in some S=1/2 honeycomb lattices with edge-sharing octahedra compounds, e.g., Na$_2$IrO$_3$, $\alpha-$RuCl$_3$ \cite{Singh2012,Banerjee2016a}.
For the S=3/2 systems, like CrXTe$_3$, the Kitaev-type exchange interaction may also exist, and recent theoretical works \cite{Xu2018} have proposed to realize the Kitaev quantum spin liquid state in CrGeTe$_3$ by applying proper in-plane strain.
As shown in the schematic of Fig.\ref{fig:SS_Models}(a), the nearest Cr-Cr pairs are proposed to have three bond-dependent Ising exchange interactions in the local $\left \{ \alpha\beta\gamma \right \} $ coordinate bases for the Kitaev model.
For simplicity, here we first discuss the Kitaev model in the perfect orthogonal $\left \{ \alpha\beta\gamma \right \} $ coordinate bases.
The Heisenberg-Kitaev Hamiltonian can be expressed as:

\begin{equation}
\label{KitaevH}
    H=-\sum_{i<j} J_{ij}  \mathbf{S}_i \cdot \mathbf{S}_j -\sum_{\left \langle i,j \right \rangle _\nu}K^{\nu}S_{i}^{\nu}S_{j}^{\nu}-A_{zz}\sum_{i} (S_{i}^z)^2 ,
\end{equation}

\noindent where the first term is the  Heisenberg exchanges term and the second term represents the nearest neighbour bond-dependent Kitaev interaction with $K^{\alpha}$=$K^{\beta}$=$K^{\gamma}$=$K$.
To obtain the final magnon dispersion, the Kitaev term in the Hamiltonian expressed by the local $\left \{ \alpha\beta\gamma \right \} $ basis can be converted into the global $\left \{ xyz \right \} $ coordinate in advance to apply the Holstein-Primakoff transformation and Fourier transformation, and then the total Hamiltonian can be rewrote in momentum space by bosonic operators.
According to the linear-spin-wave theory, we can ignore the high-order terms and just keep the ground-state energy term and the linear terms which contain the magnon relation.
By fitting the experimental results with the model, we can easily obtain the exchange parameters.
Comparing to the previous Heisenberg-DM model, the DM term is just replaced by the Kitaev term, and also up to five isotropic Heisenberg interactions are included for CrXTe$_3$.
To make the calculated magnon bands consistent with the measured band dispersion, the optimized Kitaev interaction parameters are quite large comparing to their nearest-neighbor Heisenberg interaction for both CrSiTe$_3$ and CrGeTe$_3$.
The corresponding exchange parameters for CrXTe$_3$ are listed in Table.\ref{table:Kitaev}.
As shown in Fig.\ref{fig:SS_Models_Band}, the magnon band of CrSiTe$_3$ reproduced by the Heisenberg-Kitaev model is quite similar to the result of the pure Heisenberg model but opens a gap in the vicinity of the $K$ point like that of the Heisenberg-DM model.

However, it is important to realize that the $\left \{ \alpha\beta\gamma \right \} $ basis is not perfectly orthogonal in CrXTe$_3$, the interangle between the local $\left \{ \alpha\beta\gamma \right \} $ basis vectors is about 93.6$^\circ$ and 100.1$^\circ$ for CrSiTe$_3$ and CrGeTe$_3$ respectively.
As shown in the schematic picture of Fig.\ref{fig:SS_Models}(b), the local $\left \{ \alpha\beta\gamma \right \} $ basis vectors tilt from the $z$ axis towards the $xy$ plane by an angle $\theta$, and only when $\theta=\theta_0\approx54.74^\circ$, namely $\tan \theta=\sqrt{2}$, the $\left \{ \alpha\beta\gamma \right \} $ basis becomes perfectly orthogonal\cite{Zhang2020c}.
Here in our paper, we take $\theta$ as 57.28$^\circ$ and 62.26$^\circ$ from the experimental refined atomic structures of CrSiTe$_3$ and CrGeTe$_3$ respectively.
The imperfection of $\theta$ will yield non-zero off-diagonal components when the exchange matrix is converted into the orthogonal $\left \{ \alpha^{\prime}\beta^{\prime}\gamma^{\prime} \right \} $ basis and eventually induce an extra exchange anisotropy along the global $z$ axis.
Three general Kitaev interaction matrices in a local $\left \{ \alpha\beta\gamma \right \} $ basis can be expressed as:

\begin{equation}\label{Mat_Kitaev}
\begin{pmatrix}
 K^\alpha&  & \\
 & 0 & \\
 & & 0
\end{pmatrix}
,\quad
\begin{pmatrix}
0&  & \\
 & K^\beta  & \\
 & & 0
\end{pmatrix}
,\quad
\begin{pmatrix}
 0&  & \\
 & 0 & \\
 & & K^\gamma
\end{pmatrix}
,\quad
\end{equation}

\noindent and in the orthogonal $\left \{ \alpha^{\prime}\beta^{\prime}\gamma^{\prime} \right \} $ basis, they can be rewritten as:

\begin{equation}\label{Mat_Kitaev_ortho}
\begin{pmatrix}
 A^2 & AB & AB\\
 AB & B^2 & B^2\\
 AB & B^2 & B^2
\end{pmatrix}
\cdot \frac{K^\alpha }{C^2}
,\quad
\begin{pmatrix}
B^2 & AB & B^2\\
AB & A^2  & AB\\
B^2 & AB & B^2
\end{pmatrix}
\cdot \frac{K^\beta }{C^2}
,\quad
\begin{pmatrix}
 B^2 & B^2 & AB\\
 B^2 & B^2 & AB\\
 AB & AB & A^2
\end{pmatrix}
\cdot \frac{K^\gamma }{C^2}
,\quad
\end{equation}

\noindent where $A=2\sin\theta\cos\theta_0+\cos\theta\sin\theta_0$, $B=-\sin(\theta-\theta_0)$ and $C=3\sin\theta_0\cos\theta_0$.
Especially when $\theta=\theta_0$, then we have $B=0$ and $A=C$, the Eq.\ref{Mat_Kitaev_ortho} is immediately degenerated into the simple form the same as the Eq.\ref{Mat_Kitaev}.
However for CrSiTe$_3$ and CrGeTe$_3$ here, the off-diagonal term is non-zero due to $\theta\ne \theta_0$, hence for example one of the Kitaev matrices will representatively be

\begin{equation}
\begin{pmatrix}
 0.998 & -0.032 & -0.032\\
 -0.032 & 0.001 & 0.001\\
 -0.032 & 0.001 & 0.001
\end{pmatrix}
\cdot K^{\alpha},\quad
\begin{pmatrix}
 0.983 & -0.092 & -0.092\\
 -0.092 & 0.009 & 0.009\\
 -0.092 & 0.009 & 0.009
\end{pmatrix}
\cdot K^{\alpha}.\quad
\end{equation}

\noindent In the orthogonal $\left \{ \alpha^{\prime}\beta^{\prime}\gamma^{\prime} \right \} $ basis, the general Heisenberg-Kitaev Hamiltonian can be expressed as:

\begin{eqnarray}
\label{Kitaev_H_Gerneral}
\begin{aligned}
    H=&-\sum_{i<j} \tilde{J}_{ij} \mathbf{S}_i \cdot \mathbf{S}_j -\sum_{\left \langle i,j \right \rangle _{\nu}}\tilde{K}^{\nu}S_{i}^{\nu}S_{j}^{\nu}-\sum_{\left \langle i,j \right \rangle _{\mu\nu}}\mathit{\Gamma}_1 \left (S_{i}^{\mu}S_{j}^{\nu}+S_{i}^{\nu}S_{j}^{\mu}\right )\\
    &-\sum_{\left \langle i,j \right \rangle _{\lambda\mu\nu}}\mathit{\Gamma}_2 \left (S_{i}^{\lambda}S_{j}^{\mu}+S_{i}^{\mu}S_{j}^{\lambda}+S_{i}^{\lambda}S_{j}^{\nu}+S_{i}^{\nu}S_{j}^{\lambda}\right )-A_{zz}\sum_{i} \left (S_{i}^z\right )^2\quad,
\end{aligned}
\end{eqnarray}

\noindent where $\mathit{\Gamma}_1=\frac{B^2}{C^2}K$, $\mathit{\Gamma}_2=\frac{AB}{C^2}K$, $\tilde{K}^{\nu}=\frac{(A^2-B^2)}{C^2}K$ and $\tilde{J}_{ij}=J_{ij}+\frac{B^2}{C^2}K$ for the nearest-neighbor exchange interactions, $\tilde{J}_{ij}=J_{ij}$ for others, and $(\lambda,\mu,\nu)$ is any permutation of orthogonal basis vector's indexes $(\alpha^\prime,\beta^\prime,\gamma^\prime)$.

Although the non-zero off-diagonal terms in the matrix looks small, but here they will create an quite large negative component (e.g. CrSiTe$_3$: -0.7 meV, CrGeTe$_3$: -5.7meV ) on the magnon gap at the Brillouin center and make the energy eigenvalues become imaginary unless an extra considerable anisotropy can be introduced to cancel it out.
If so, either the anisotropy of the exchange interaction or the single ion anisotropy should be very strong.
However, as we known from the magnetization and the critical behavior, the magnetic anisotropy in CrXTe$_3$ is quite small actually.
Even if the magnon dispersion can be reproduced very well when the bond-dependent Kitaev interactions are assumed to be perfectly orthogonal, it is still difficult to understand the huge difference of the strength of the Kitaev interaction between CrSiTe$_3$ and CrGeTe$_3$, since the Kitaev interaction mainly arise from the heavy ligands (namely, Te of CrXTe$_3$) \cite{Xu2018,Xu2020}.
We can not completely exclude the existence of the Kitaev interaction, but comparing to the previous DM model, the Heisenberg-Kitaev model is unlikely a proper model to describe the exchange interactions in this system  at least.

\begin{table*}[b!]
\setlength{\tabcolsep}{10mm}{
\begin{tabular}{c@{~~~}c@{~~~}c@{~~~}c@{~~~}c@{~~~}c@{~~~}c@{~~~}c}
    \toprule
        & $J_{ab1}$ & $J_{ab2}$  & $J_{c1}$ & $J_{c2}$ & $K$ \\
    \midrule
        CrSiTe$_3$ & 0.4 & 0.2 & 0.08 & 0.065 & 3  \\
    \midrule
        CrGeTe$_3$ & 0.24 & 0.42 & 0.1 & 0.08 & 6.5  \\
    \bottomrule
\end{tabular}}
\caption{
The parameters of the Heisenberg exchange interactions and the Kitaev interactions are listed together for CrSiTe$_3$ and CrGeTe$_3$.
The local $\left \{ \alpha\beta\gamma \right \} $ basis is assumed perfectly orthogonal and the single-ion anisotropy is fixed to 0.01 meV.
The unit of the parameters is meV.
}
\label{table:Kitaev}
\end{table*}

\section{Magnon band topology}
\subsection{Berry curvature and Chern number}

Formulating semiclassical equations of motion for magnon wave packets \cite{Mook2014b}, which include the anomalous velocity leads to a non-zero Berry curvature. We generally define the Berry curvature field associated with the $n^{th}$ magnon band as

\begin{equation}
 \boldsymbol{\Omega}_{n\mathbf k}= i\! \sum_{m\neq n}\!
 \frac{\langle\Psi_{n\mathbf{k}}| \partial_{\mathbf{k}} H(\mathbf{k})|\Psi_{m\mathbf{k}}\rangle \times \langle\Psi_{m\mathbf{k}} |\partial_{\mathbf{k}}H(\mathbf{k}) |\Psi_{n\mathbf{k}}\rangle }{(\epsilon_{n\mathbf{k}}-\epsilon_{m\mathbf{k}})^2},
\end{equation}
 
\noindent where $\partial_{\mathbf k} H(\mathbf k)$ denotes the gradient of the Hamiltonian in momentum space.
$\epsilon_{m\mathbf{k}}$ and $\epsilon_{n\mathbf{k}}$ are the magnonic eigenvalues.

Based on this Berry curvature, we can classify the topology of the $n^{th}$ magnon branch using  Chern number:

\begin{equation}
 C(P)=\frac{1}{2\pi}\int_{P}\boldsymbol{\mathrm{\Omega}}(\mathbf{k})\cdot\mathbf {\vec{n}} \,dP,
\end{equation}

\noindent where $P$ is a two-dimensional slice of the BZ and $\mathbf {\vec{n}}$ is its normal vector.
In our calculation, the $\mathbf{\vec{n}}$ is selected perpendicular to the $xy$ plane.
As the lowest three branches cross with each other, the lower three branches are sum together.

\begin{figure*}
\centering
\includegraphics[width=8cm]{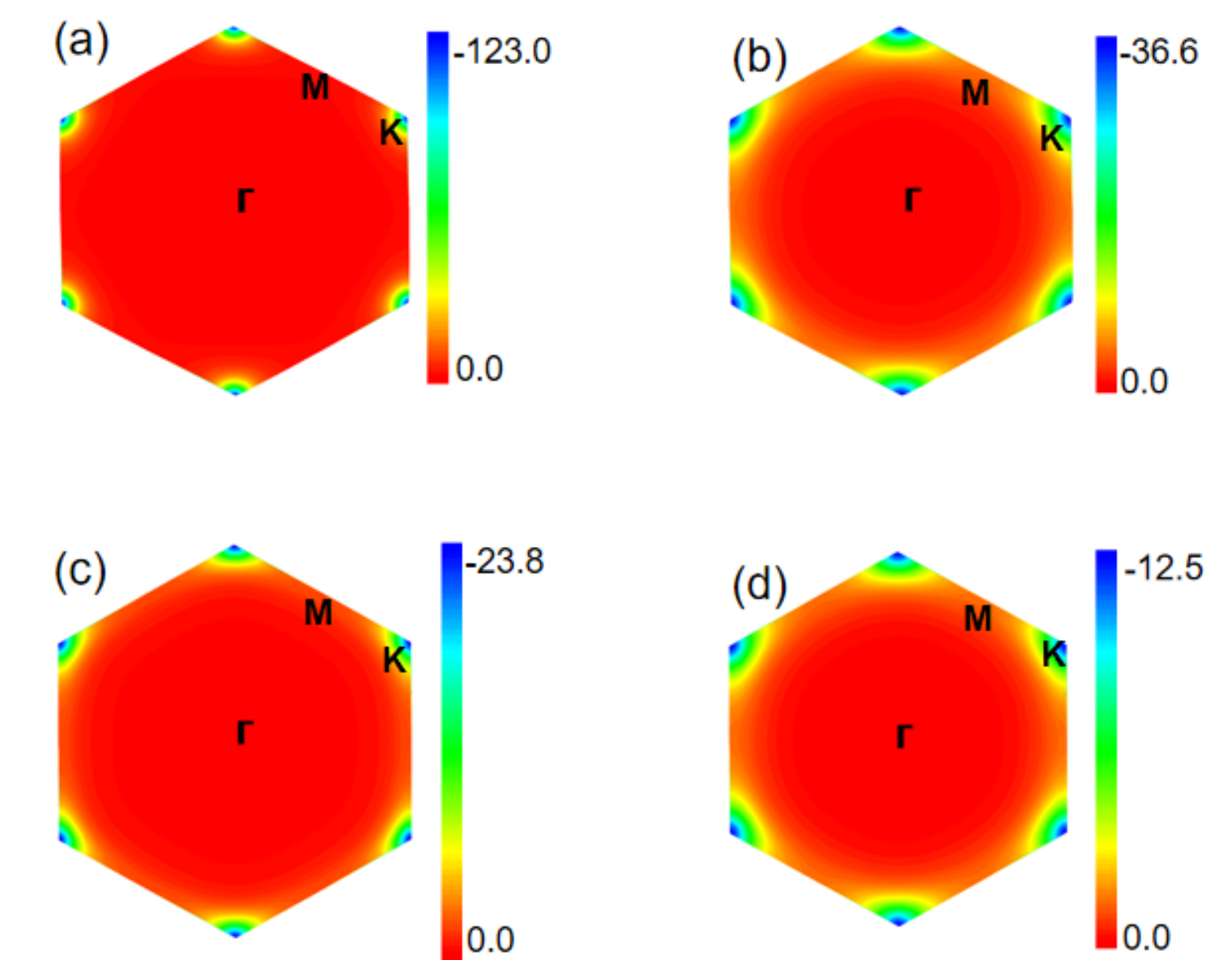}
\caption{\label{fig:SS_Berry}
\textbf{The calculated Berry curvature in the $K_x-K_y$ plane}. The Berry curvature for the lowest three magnon branches of bulk CrSiTe$_3$ (a) and CrGeTe$_3$ (b).
The Berry curvature for the lowest branch of monolayer CrSiTe$_3$ (c) and CrGeTe$_3$ (d) in the first Brillouin zone.
The Chern number is -3 for (a,b) and -1 for (c,d).
}
\end{figure*}

The Berry curvature distribution of bulk CrGeTe$_3$ in the first BZ is shown in Fig.\ref{fig:SS_Berry}(a,b).
The Chern number of the lowest three branches is $-3$ and the Chern number of the highest three branches is $+3$.
If we ignore the interlayer interaction, we can easily use the monolayer system to characterize the topological properties of the system, and the corresponding Berry curvature is shown in Fig.\ref{fig:SS_Berry}(c,d) with the Chern number $-1$ for the lowest branches.
From the figure, almost the same Berry curvature distribution can be  obtained.
Compared to the intralayer interaction, the interlayer interaction is weaker and the DM interaction only exist among intralayer.
In CrXTe$_3$, each layer contributes one optical and acoustic branches.
The interlayer interaction splits the three degenerate bands but influence little on their topological properties.

\subsection{Edge states and thermal Hall conductivity}
To simplify the calculation, the monolayer CrXTe$_3$ is utilized to study the edge states.
Here, the color scale of the bands in the main text Fig.4 is calculated based on the following equation:

\begin{equation}
 \label{eq:local}
 LW(\mathbf k,j)=\sum_i\phi^{i*}(\mathbf k,j)\phi^{i}(\mathbf k,j)(R_z^i-0.5),
\end{equation}

\noindent where $\mathbf k$ is the reciprocal space vector, $j$ denotes the band index,  $i$ numbers the magnetic atom, and $R^i_z$ represents the normalized position for atom $i$ along the $z$-axis. $\phi^i(k,j)$ is the components of the  right eigenstates of $j$ at the magnetic atom $i$.

Given the nontrivial topology, the topological magnon edge states can contribute to the transverse thermal Hall voltage under an applied longitudinal temperature gradient \cite{Onose2010a,Matsumoto2011a,Mook2014b,Owerre2016b,Aguilera2020}, namely the topological thermal Hall effect.

The energy-dependent contribution to the $i{\kern-.5pt}j$'th Cartesian component of the thermal Hall conductivity tensor $\hat{\kappa}$ can be calculated as

\begin{equation}
 \kappa^{ij}(\epsilon)=\frac{k_\text{B}^2 T}{(2\pi)^3 \hbar} \sum_n\int_\text{BZ}\delta(\epsilon_{n\mathbf{k}}-\epsilon)\,C_2 (f_n^\text{B})\, \Omega^{ij}_n(\mathbf{k})\,d\mathbf{k},
 \label{equ2}
\end{equation}

\noindent where $n$ enumerates the magnon bands, $f_n^\text{B}$ is the Bose-Einstein distribution function, which can be expressed as $f_n^\text{B}=(e^{\epsilon_{n\mathbf{k}}/k_\text{B}T}-1)^{-1}$, and $C_2$ is given by

\begin{equation}
C_2(x)=(1+x)\left(\ln\frac{1+x}{x}\right)^2-\ln^2 x-2\mathrm{Li}_2 (-x),
\end{equation}

\noindent with  $\rm{\ Li}_2$ denoting the dilogarithm function. In our calculation, only the result of transverse thermal Hall conductivity $\kappa^{xy}$ is shown, as the $\kappa^{xz}$ and $\kappa^{yz}$ is zero. The transverse thermal Hall conductivity  of the system is then defined as
$ \kappa^{xy}=\lim_{\mu\rightarrow\infty}\kappa^{xy}_{\mu}$, where $\kappa^{xy}_{\mu}=\int_0^{\mu}\kappa^{xy}(\epsilon)\,d\varepsilon$ is the cumulative thermal Hall conductivity.

From the experimental results, we know that the respective Curie temperature of CrSiTe$_3$ and CrGeTe$_3$ are around 30 K and 60 K.
The calculated temperature-dependent and energy-dependent thermal Hall conductivity are shown in Fig.\ref{fig:SS_thermalHall}.
For both materials, $\kappa^{xy}$ is significant enhancement in the energy region close to the band gap, which can be attributed to the distribution of the Berry curvature around the $K$ point.
In low temperature only these "topologically-trivial" states are excited according to the Bose-Einstein distribution, leading to the zero platform at very low temperature.
Then conductivity increases as the temperature increases.
The effect of the DM interaction is shown in Fig.\ref{fig:SS_thermalHall}(e), from which we can observe that the thermal Hall conductivity increases with the enhancement of the DM Interaction.

The predicted thermal Hall conductivity for CrGeTe$_3$ reaches the order of $10^{-4}$ W/Km, which is large enough to be observed in experiment.
Furthermore, the DM interaction value is determined by the strength of spin-orbit coupling.
Compared to CrGeTe$_3$ and CrSiTe$_3$, CrSnTe$_3$ and CrPbTe$_3$ should have even stronger spin-orbit coupling and some recent theoretical studies have demonstrated that the corresponding monolayer system could have a ferromagnetic state with higher Curie temperatures \cite{Zhuang2015,Khan2020}.
These materials are thus more likely to have even bigger thermal Hall conductivity than Lu$_2$V$_2$O$_7$ \cite{Katsura2010,Onose2010a}.

\cleardoublepage

\begin{figure*}
\centering
\includegraphics[width=12cm]{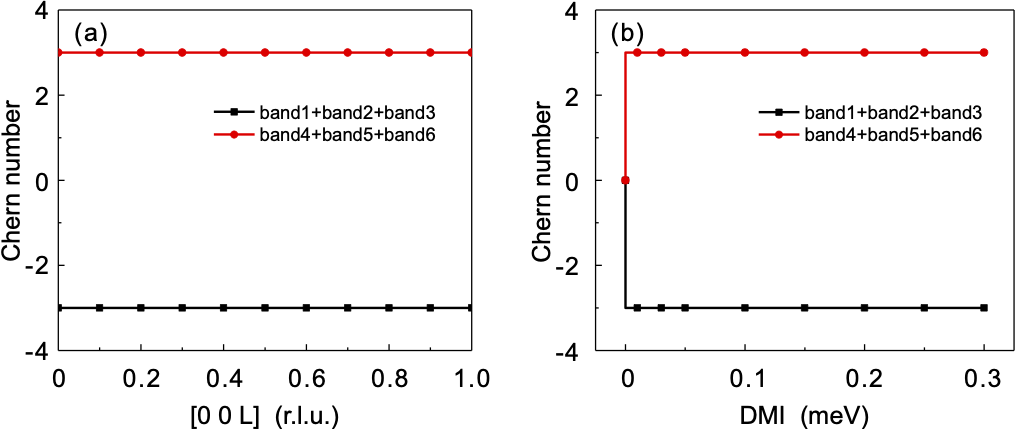}
\caption{\label{fig:SS_nontrival}
\textbf{Chern numbers of magnon bands}.
(a) The Chern number along the [0,0,L] direction.
The sum of the Chern numbers are -3 and 3 for the lowest and highest three bands respectively.
(b) The influence of the magnitude of the DM interaction on the Chern numbers of the bands.
}
\end{figure*}

\begin{figure*}
\centering
\includegraphics[width=12cm]{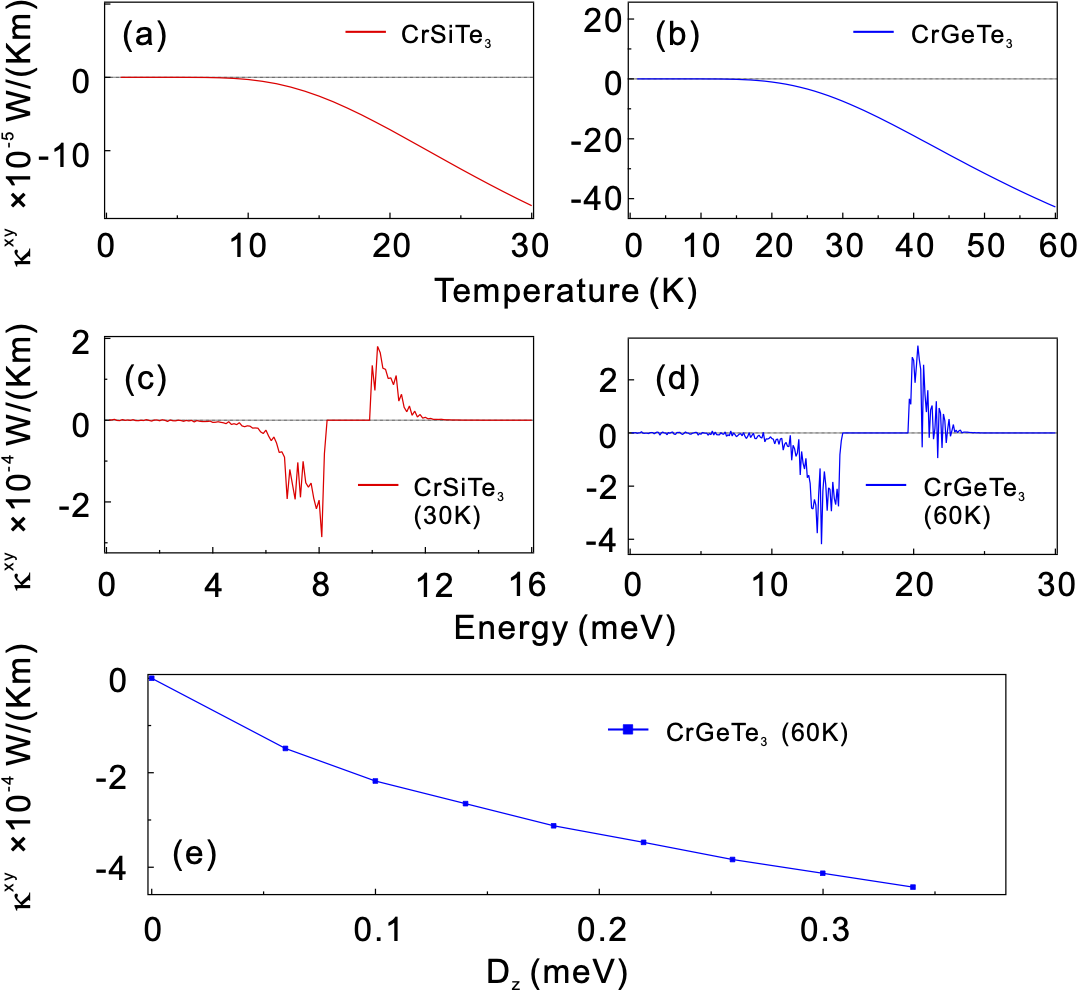}
\caption{\label{fig:SS_thermalHall}
\textbf{The calculated transverse thermal Hall conductivity ($\kappa^{xy}$) of CrGeTe$_3$ and CrSiTe$_3$}.
The temperature dependence (a, b) and energy dependence (c, d) of the thermal Hall conductivity of CrGeTe$_3$ and CrSiTe$_3$.
The $\kappa^{xy}$ with the function of $D_z$ for CrGeTe$_3$ at 60K is shown in (e).
}
\end{figure*}

\end{document}